\documentclass[twocolumn,pra,superscriptaddress]{revtex4-1}

\usepackage{epsfig}
\usepackage{graphicx}
\usepackage{amsmath}
\usepackage{amssymb}
\usepackage{epstopdf}
\usepackage{xcolor}
\usepackage{subfigure}
\usepackage{bm}
\usepackage{dsfont}
\usepackage{hyperref}
\usepackage{braket}
\usepackage{todonotes}
\usepackage{physics}
\usepackage[normalem]{ulem}

\newcommand{\mi}{\mathrm{i}}

\begin{document}

\title{Storing vector-vortex states of light in intra-atomic frequency comb}

\author{Chanchal}
\affiliation{Indian Institute of Science Education and Research, Mohali, Punjab 140306, India}

\author{ G.~P.~Teja} 
\affiliation{Indian Institute of Science Education and Research, Mohali, Punjab 140306, India}

\author{ Christoph Simon} 
\affiliation{Institute for Quantum Science and Technology and Department of Physics and Astronomy, University of Calgary, Canada T2N~1N4}

\author{Sandeep K.~Goyal}
\email{skgoyal@iisermohali.ac.in}
\affiliation{Indian Institute of Science Education and Research, Mohali, Punjab 140306, India}

\begin{abstract}
  Photons are one of the prominent candidates for long-distance quantum communication and quantum information processing. Certain quantum information processing tasks require storage and faithful retrieval of single photons preserving the internal states of the photons. 
  Here we propose a method to store the vector-vortex states of light in the intra-atomic frequency comb based quantum memory. We show that an atomic ensemble with two intra-atomic frequency combs corresponding to $\Delta m = \pm1$ transitions of similar frequency are sufficient for a robust and efficient quantum memory for vector-vortex states of light. As an example, we show that the Cs and Rb atoms are good candidates for storing these internal modes of light.

\end{abstract}

\maketitle

\section{Introduction}

Efficient  quantum computation and quantum information processing require  quantum systems with long coherence time and high degree of control. Although, there are several suitable candidates such as Nuclear spins, quantum dots, superconducting qubits, and trapped ions, photons are  among the strongest candidates for long-distance quanutm communication and quantum computation \cite{flamini2018}. Linear optical quantum computation \cite{kok2007linear}, quantum key distribution \cite{pirandola2020}, quantum teleportation \cite{bouwmeester1997,wang2015}, quantum repeaters \cite{sangouard2011} are few of the examples where photons have shown dominance. The choice of photons for quantum communication is natural as they can travel great distances without much trouble.

Some of the most notable degrees of freedom of photons that are used for QIP tasks are polarization, time-bins and the orbital angular momentum \cite{flamini2018}. While the polarization space is two-dimensional, the time-bins and orbital angular momentum space is potentially infinite dimensional, which enables high information carrying capacity in individual photons. One of the biggest challenge in photonic quantum information processing tasks is to store and retrieve the photons while preserving their internal states in an efficient and controllable way. There are several such protocols \cite{simon2010,heshami2016}, e.g., electromagnetically induced transparency \cite{fleischhauer2002,hsiao2018,wang2019}, controlled reversible inhomogeneous broadening \cite{nilsson2005,kraus2006,iakoupov2013}, atomic frequency combs \cite{afc2009,afzelius2010,businger2020}, intra-atomic frequency combs \cite{iafc19}, gradient echo memory \cite{hetet2008,hosseini2011,sparkes2013}, and Raman memory \cite{nunn2007,reim2011,guo2019}. In all these protocols, a photon is made to interact with an ensemble of atoms or atom-like systems, carefully tuned to maximize the absorption of the photons. A controlled sequence of pulses can switch on and off the interaction between the photon and the atomic ensemble, hence resulting in a controlled storage. 

Typically, in an atomic ensemble based quantum memories, the atoms are tuned to interact with a single polarization. Therefore, one can not store polarization states of light in such systems. To overcome this problem a number of solutions have been implemented. For example, in EIT based quantum memories the orthogonal polarization states of the input light are mapped to  two distinguished paths with the same polarizations and absorbed in the atomic ensemble. Finally, when the light is retrieved the paths are mapped to the polarization states at the output~\cite{vernaz2018,wang2019,t2015,namazi2017,xu2013}. A similar technique is used in AFC based quantum memory to store polarization DoF of light~\cite{afc2012,afcsw2012,prl2012,laplane2015}. The  EIT and AFC based quantum memories have been shown to store transverse modes \citep{prl2004,prl2007,pra2009,ding2013,oam2013,nicolas2014,bai2017,pra2017,yang2018,pra2020}; however only EIT is extended to store polarization and OAM modes simultaneously \cite{nat2015,ye2019t}. However, mapping one DoF onto other  in order to preserve it in the storage process is not an efficient way to store such DoFs. 

Apart form these two techniques, an atom-cavity system is used to implement polarization storage \cite{specht2011,kalb2015}, nevertheless the cavity protocols have the drawback of trapping single atoms and requirement of feed back pulses for the deterministic storage \cite{kalb2015} or optimization of control pulses for storage and retrieval of polarization qubits \cite{specht2011}.

A CRIB protocol with two orthogonal transitions is also proposed to store polarization \cite{crib2011,cribpra2011}
which requires reversing of detunings in a controlled fashion and applying a position dependent phase for an efficient retrieval.

Another protocol for quantum memory based on Intra-atomic frequency comb (I-AFC) has been shown to be robust and efficient~\cite{iafc19,teja2021}. In an I-AFC, the frequency comb is constructed from the transitions between hyperfine energy levels from individual atoms. Since each of the atom contains a frequency comb, I-AFC based quantum memory is robust against phase fluctuations and uniformity in the comb structure~\cite{teja2021}.

Here we propose a scheme to store the vector-vortex states of light using intra-atomic frequency combs (I-AFC). Vector-vortex states are the quasi-entangled states between polarization and OAM DoFs of light and are very useful for QIP tasks and quantum metrology~\cite{barreiro2010,cozzolino2019,d2013photonic}. We show that I-AFC is a natural candidate to store vector-vortex states by showing that it can store individually the polarization and OAM modes efficiently. Unlike EIT based quantum memory, I-AFC doesn't require high optical-depths and elongated atomic traps to store vector vortex states of light.
I-AFC can be easily realized by Zeeman splitting the hyperfine levels in atoms and possesses all the necessary features of a typical AFC \cite{iafc19}. This makes I-AFC a feasible tool to implement protocols using OAM and polarization qubits. We also show that I-AFC in the Cs and Rb atoms can be employed to store vector-vortex states.

The article is organised as follows: we start with the relevant background in in Sec.~\ref{Sec:Background} where we discuss the I-AFC based quantum memory, LG modes, vector-vortex beams, and discuss the effects of the Doppler shift on the LG modes. In the Sec.~\ref{Sec:Results}, we present the results on storing vector-vortex states of light in ideal I-AFC systems. We also discuss factors which might affect the quality of the storage. Finally we show numerically that an ensemble of Cs and Rb atoms is capable of storing Vector-vortex states efficiently under appropriate conditions.
We conclude in Sec.~\ref{Sec:Conclusion}.

\section{Background}\label{Sec:Background}
In this section, we introduce the concepts and techniques relevant for our results. We start with I-AFC based quantum memory. We also discuss the LG-modes of light and vector-vortex beams, and the effect of Doppler shift on the LG-modes. 

\subsection{I-AFC based quantum memory}\label{scheme}
\begin{figure}[!htb]
\subfigure[\label{comb}]{\includegraphics[height=3cm,width=4.4cm]{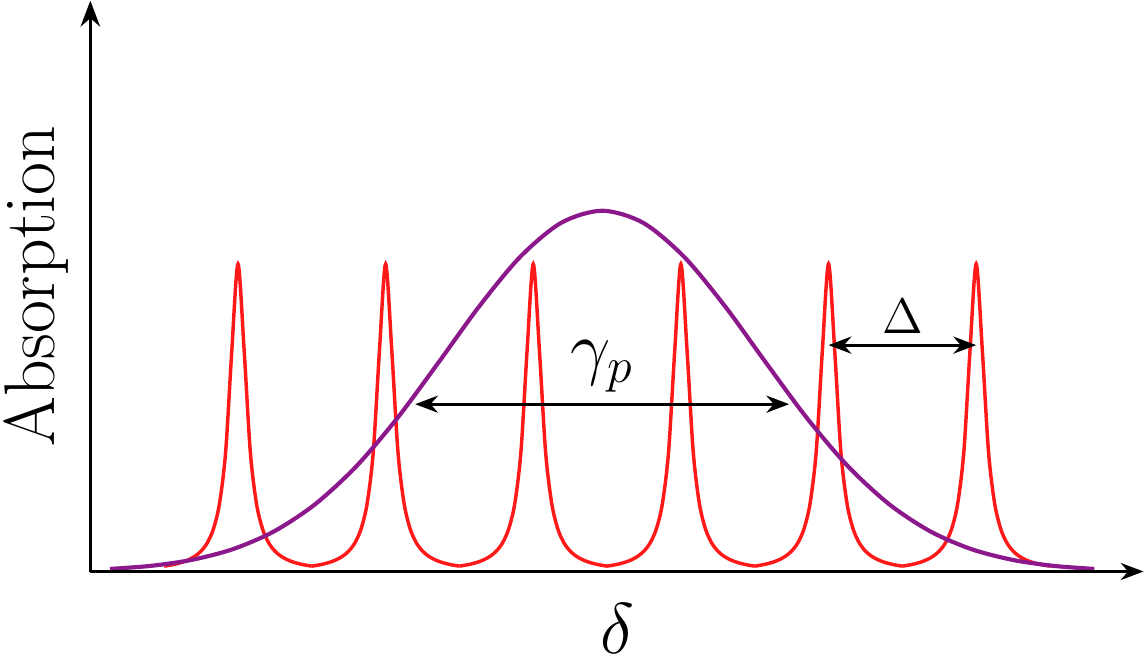}}
\subfigure[\label{echo}]{\includegraphics[height=3cm,width=4.1cm]{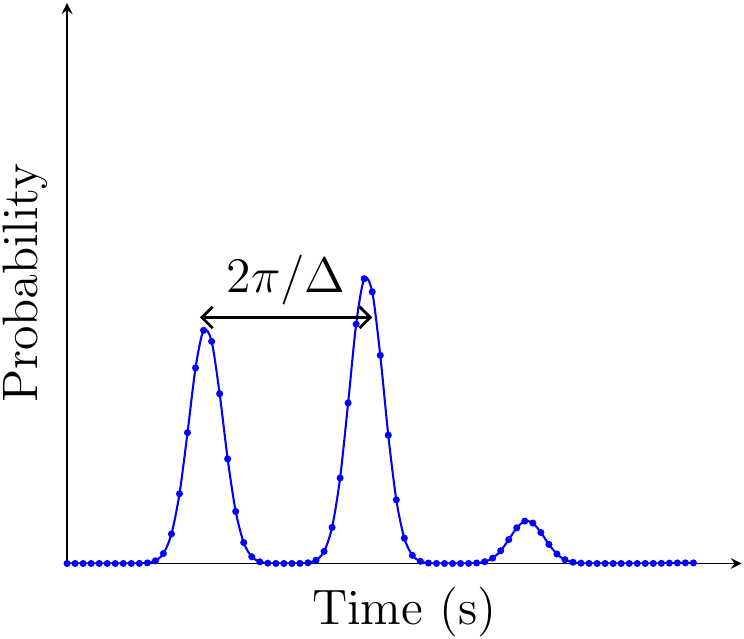}}
\caption{\subref{comb} I-AFC with comb spacing $\Delta$ interacting with input pulse of width $\gamma_p$. \subref{echo} Typical photon-echo after time a time delay of $2\pi/\Delta$.}
\end{figure}

The I-AFC is a frequency comb constructed from the dipole allowed transitions between the hyperfine levels of the ground state and the excited state of an atom. The degeneracy in the hyper-fine levels of the ground and excited states is lifted by applying an  external magnetic field. All these transitions between the ground level and the excited level collectively result in I-AFC,~Fig.~\ref{comb}.

For an ensemble of atoms with each atom constituting a frequency comb, if a weak electromagnetic pulse $\mathcal{E}(z,t)$ with spectral width $\gamma_p$ is passed through it, the dynamics of the state $\rho(z,t)$ of the atomic ensemble and the electromagnetic field amplitude $\mathcal{E}(z,t)$  can be written as~\cite{iafc19}
\begin{align}
\qty(\pdv{z}+\dfrac{1}{c}\pdv{t}){\mathcal{E}}(z,t)&=\dfrac{\mi\omega_L}{2\epsilon_0c}{\mathcal{P}}(z,t),\label{e11}\\ 
\pdv{\rho_{nm}(z,t)}{t}+\qty(\mi\Delta_{nm}+\dfrac{\gamma}{2})\rho_{nm}(z,t)&=\mi \dfrac{d_{nm}\mathcal{E}(z,t)}{2 \hbar}\rho_{mm}.\label{e22}
\end{align}
Here   $\Delta_{nm}$ is the detuning between the $\ket{e_n}\leftrightarrow\ket{g_m}$ transition and the mean frequency of light $\omega_L$. $\gamma$ is the natural line-width of the atomic transitions which we are considering to be the same.  The initial population in the ground state $\ket{g_m}$ is $\rho_{mm}$, whereas $\rho_{nm} \equiv \bra{e_n}\rho\ket{g_m}$ is the matrix element of the state $\rho$.  $d_{nm}$ is the transition dipole moment between $\ket{e_n}\leftrightarrow\ket{g_m}$. Since a typical dipole allowed atomic transition absorbs and emits light in accordance with the transition selection rules $\Delta m=0,\pm 1$, we write $d_{nm}$ in the spherical basis, hence the dipole matrix element $d_{nm}$ is always real~\cite{bransden2003}.


The induced polarization $\mathcal{P}$ of the atomic ensemble can be written as a function of the atomic state $\rho$ as
\begin{align}
\mathcal{P}(z,t)=2\mathcal{N}\sum_{n,m}d_{nm}^*\rho_{nm},\label{P}
\end{align}
where $\mathcal{N}$ is the atomic number density. Since $d_{nm}$ is real, the $*$ from Eq.\eqref{P} can be omitted.

On solving Eq.~(\ref{e11}) and (\ref{e22}), one gets the following expression for the output electric field in the frequency domain \cite{iafc19}
\begin{align}
\tilde{{\mathcal{E}}}(z,\omega)=&e^{-\mathcal{D} z} e^{-\mi\omega z/c} \tilde{{\mathcal{E}}}(0,\omega),\label{e13}
\end{align}
where $\tilde{{\mathcal{E}}}(0,\omega)$ is the input electric field amplitude and $\mathcal{D}$ is given by
\begin{align}
\mathcal{D}=\sum_{n,m}\dfrac{g_{mm}}{\qty[\mi(\Delta_{nm}+\omega)+\dfrac{\gamma}{2}]}{d^2_{nm}},
g_{mm}=\dfrac{\omega_L\mathcal{N}\rho_{mm}}{2c\hbar\epsilon_0}.\label{pr}
\end{align}

In the ideal case, when the comb spacing $\Delta_{nm}$ and the dipole matrix elements $d_{nm}$ are same for all the neighbouring transitions, a photon-echo is observed in the output at times which are the multiples of $2\pi/\Delta$ (Fig.~\ref{echo}) and negligible light is emitted between the echos. In non-ideal cases, when the frequency comb is non-uniform, the photon-echo may be observed at $2\pi/\Delta'$, for some  effective comb spacing  $\Delta'$~\cite{teja2021} with lower efficiency. In this way, the I-AFC system behaves like a delay line. To achieve on-demand quantum memory, the excitation is transferred from the excited level to a long-lived spin level by applying an appropriate $\pi$-pulse. Another such pulse will transfer the excitation back to the excited level which will cause the photon-echo.

The quality of the quantum memory can be expressed in terms of two parameters-- storage efficiency $\eta$ and the fidelity $\mathcal{F}$ between the input and output states of light. The storage efficiency of the quantum memory in the I-AFC protocol is defined as the ratio of the output intensity in the first echo to the total input intensity of light and reads \cite{iafc19}
\begin{align} 
\eta=\dfrac{\int_{\pi/\Delta}^{3\pi/\Delta}dt \abs{\mathcal{E}(z = L,t)}^2}{\int\,dt \abs{\mathcal{E}(z = 0,t)}^2 },\label{eff}
\end{align}
where $L$ is the length of the atomic ensemble.

The fidelity of the quantum memory describes the amount of overlap between the input state $\ket{\Psi_{in}}$ and the output state $\ket{\Psi_{out}}$, and can be formally written as
\begin{align}
\mathcal{F}=\abs{\ip{{\Psi_{\text{in}}}}{\Psi_{\text{out}}}}^2.
\end{align}
Since in I-AFC, the output electric field comes out at time $2\pi /\Delta$, the fidelity in the I-AFC scheme between the input electric field $\vb{\mathcal{E}}_{\text{in}}(t)$ and the first-echo is given as

\begin{align}
\mathcal{F}=&\dfrac{\abs{\int_{\pi/\Delta}^{3\pi/\Delta}dt~\braket{\vb*{\mathcal{E}}_{\text{in}}(t- 2\pi/\Delta)}{\vb*{\mathcal{E}}_{\text{out}}(t)}  }^2}{\qty[\int dt~\braket{\vb*{\mathcal{E}}_{\text{in}}}]\qty[\int dt ~\braket{\vb*{\mathcal{E}}_{\text{out}}}]}.\label{fid}
\end{align}

The most general state of light including the polarization and the transverse profile can be written as
\begin{align}
\boldsymbol{\mathcal{E}}(t)=\mqty[{{\mathcal{E}}_{+}(x,y,t) \\ {\mathcal{E}}_{-}(x,y,t)}],
\end{align}
where $\mathcal{E}_{\pm}$ corresponds to the right(left) polarization component of the electric field and $x,y$ are the transverse coordinates. In such cases, the expression for the fidelity between input and output can be written as
\begin{widetext}
\begin{align}
\mathcal{F}=
\dfrac{\abs{\int_{\pi/\Delta}^{3\pi/\Delta}dt~ \int dx\, dy~\big[{\mathcal{E}}^*_{\text{in}+}(x,y,t- 2\pi/\Delta){\mathcal{E}}_{\text{out}+}(x,y,t)+{\mathcal{E}}^*_{\text{in}-}(x,y,t- 2\pi/\Delta){\mathcal{E}}_{\text{out}-}(x,y,t)\big]}^2}{\qty(\int dt~ \int dx\, dy~(\abs{\mathcal{E}_{\text{in}+}}^2+\abs{\mathcal{E}_{\text{in}-}}^2))\qty(\int dt~ \int dx\, dy~(\abs{\mathcal{E}_{\text{out}+}}^2+\abs{\mathcal{E}_{\text{out}-}}^2))}\label{fidp}.
\end{align}
\end{widetext}



\subsection{LG modes and Vector-vortex beams}
Laguerre-Gauss (LG) modes are the eigenmodes of the paraxial wave-equation~\cite{allen92,allen99}. They are also the eigenmodes of angular momentum operators. Hence, the LG beams possess certain orbital angular momentum. In the cylindrical coordinates, the expression for the LG modes read
\begin{align}\label{v1}
\begin{aligned}
LG^\ell_p(r,\phi,z)&=\dfrac{C}{w(z)}\qty(\dfrac{\sqrt{2}r}{w(z)})^{|\ell|}
L^{|\ell|}_p \qty(\dfrac{2 r^2}{w(z)^2}) \exp(\dfrac{-r^2}{w(z)^2})\\
&\times\exp(\dfrac{\mi kr^2}{2\bar z})\exp[-\mi(2p+\abs{\ell}+1)\psi(z)]\exp(\mi\ell\phi)\\
\equiv& f_{\ell}^p(r,z)\exp(\mi\ell\phi),&
\end{aligned}
\end{align}
where
\begin{align}
\begin{aligned}
w(z)=w_{0}\sqrt{1+\qty(\dfrac{z}{z_{R}})^2},~\bar{z}=\dfrac{z^2+z_{R}^2}{z},
\end{aligned}
\end{align}
and $\psi(z)=\tan^{-1}\qty(\dfrac{z}{z_{R}})$ is the Guoy phase. Here $\omega_0$ is the beam waist at $z=0$, $z_{R}=\pi w_{0}^2/\lambda$ represents the Rayleigh range, $C$ is the normalization constant, $L^{|\ell|}_p$ is the associated Laguerre polynomial, $p\geq 0$ is the radial index and $-\infty < \ell < \infty$ is the azimuthal index.  $\ell \hbar$ is the orbital angular momentum (OAM) per photon for a given LG mode.

Transverse LG modes along with the polarization will give the general state of the paraxial light which can be written as
\begin{align}
\boldsymbol{\mathcal{E}}(\vb{r_{\perp}})=
\sum_{\ell,p}\left(\alpha_{\ell,p} LG_p^\ell(\vb{r_{\perp}})\ket{R}+\beta_{\ell,p} LG_{p}^{\ell}\ket{L}\right),\label{lgm}
\end{align}
Here $\ket{R}$ ($\ket{L}$) corresponds to the right (left) circular polarization and $\alpha_{\ell,p} ,\beta_{\ell,p} \in \mathds{C}$ such that $\sum_{\ell,p}\qty(\abs{\alpha_{\ell,p}}^2+\abs{\beta_{\ell,p}}^2)=1$.

\begin{figure}[!htb]\label{vortex}
\subfigure[\label{vvr}]{\includegraphics[height=4cm,width=4cm]{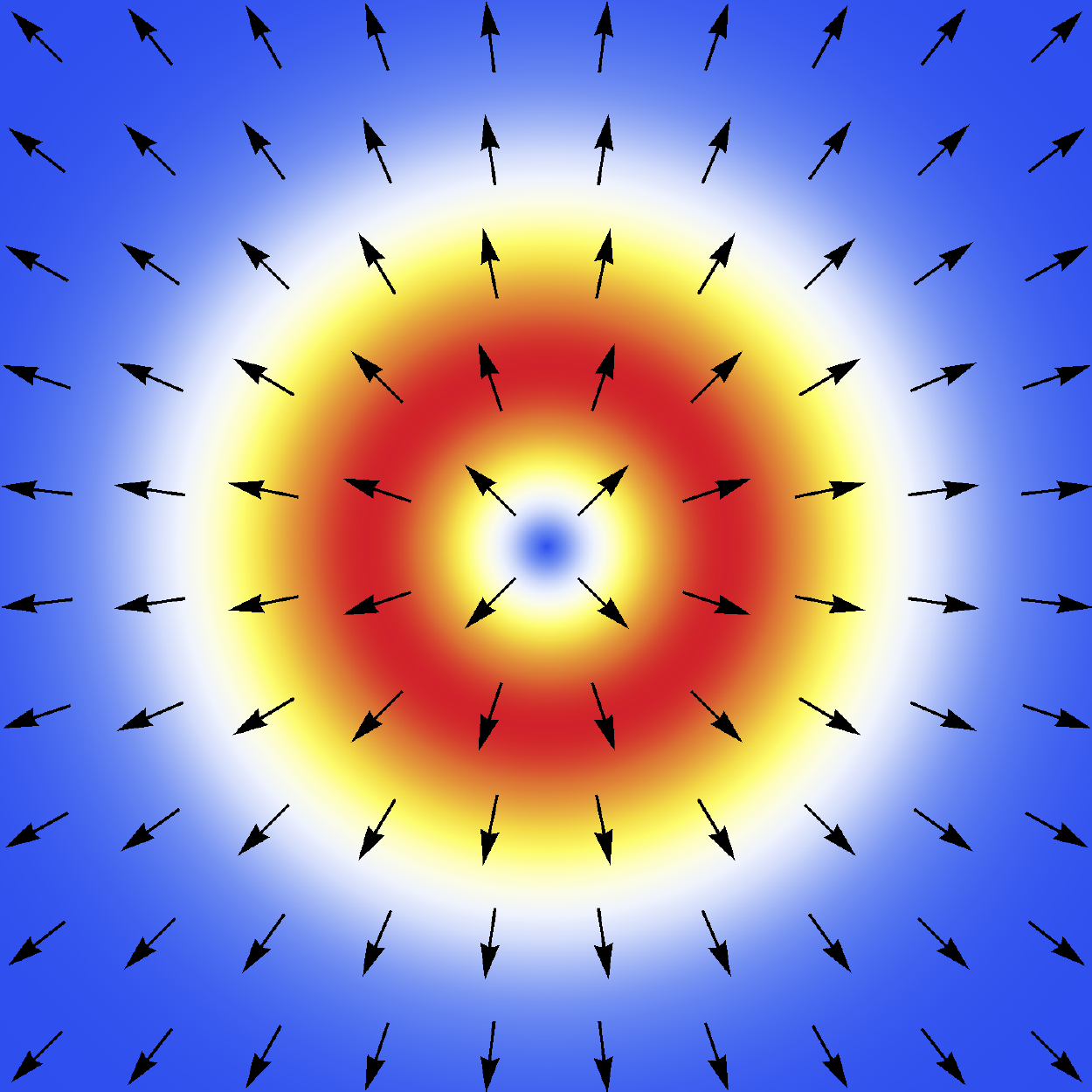}}
\subfigure[\label{vva}]{\includegraphics[height=4cm,width=4cm]{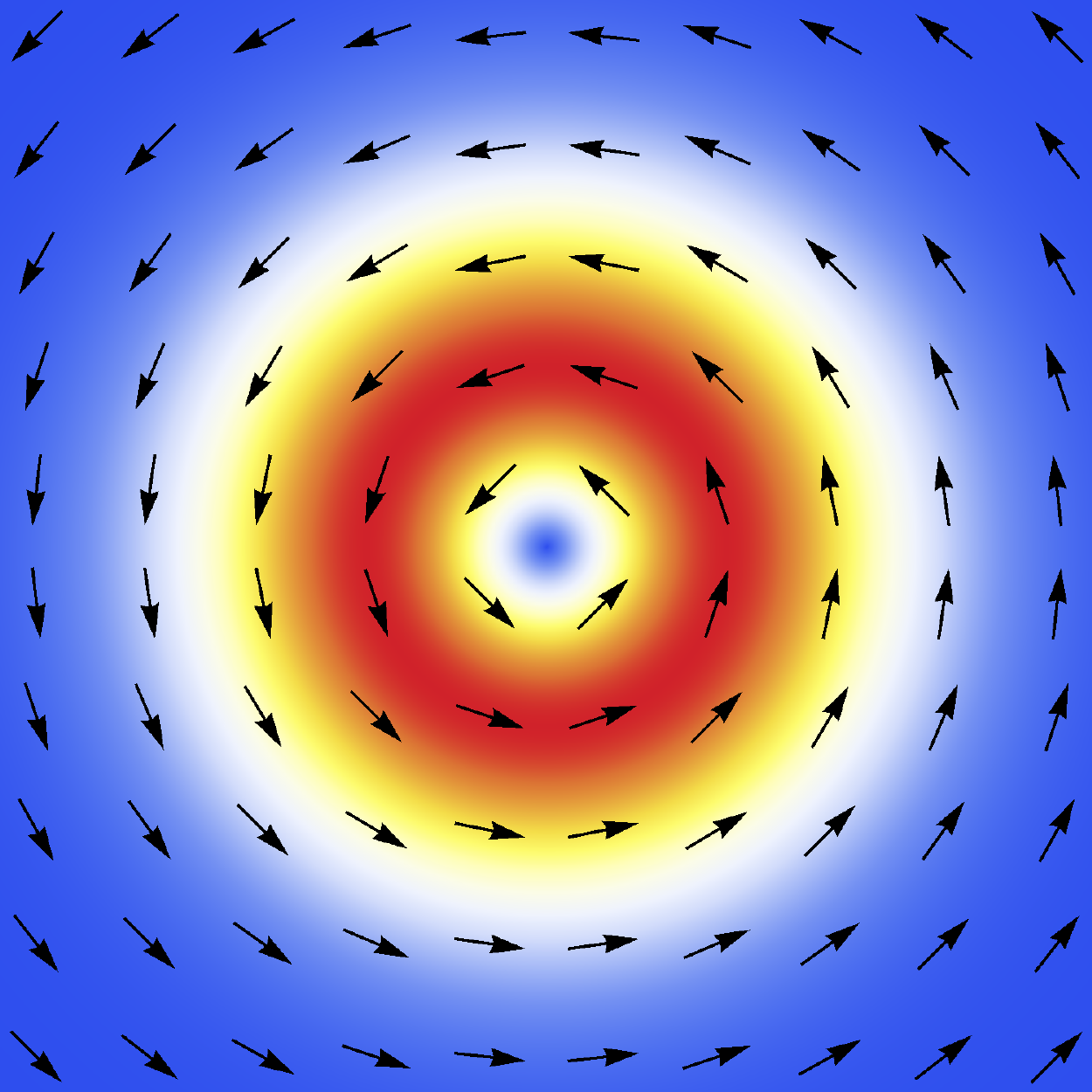}}
\caption{Polarization distribution in Vectors-Vortex states. \subref{vvr} \subref{vva} corresponds to the states 
$\dfrac{1}{\sqrt{2}}\qty[\ket{\ell}\ket{R} + \ket{-\ell}\ket{L}]$ and $\dfrac{\mi}{\sqrt{2}}\qty[\ket{\ell}\ket{R} - \ket{-\ell}\ket{L}]$ respectively.}
\end{figure}

In this work, we set $p=0$ and represent $LG^\ell_0$ by $\ket{\ell}$. Furthermore, we keep to only $\pm \ell$ values of OAM and consider the states of the form
\begin{align}
\boldsymbol{\mathcal{E}}(\vb{r_{\perp}})=\qty[\alpha\ket{\ell}\ket{R} + \beta\ket{-\ell}\ket{L}].\label{vv}
\end{align}
These states are called vector-vortex beams \cite{pra16,nat2015} and they exhibit a location position polarization in the transverse plane. In Fig.~\ref{vortex} we show two such vector-vortex states for different choices of $\alpha$ and $\beta$.

\subsection{Effect of Doppler shift on OAM states of light}\label{dp} 

 \begin{figure*}[!tbh]
   \subfigure[\label{idl1}]{\includegraphics[scale=0.35]{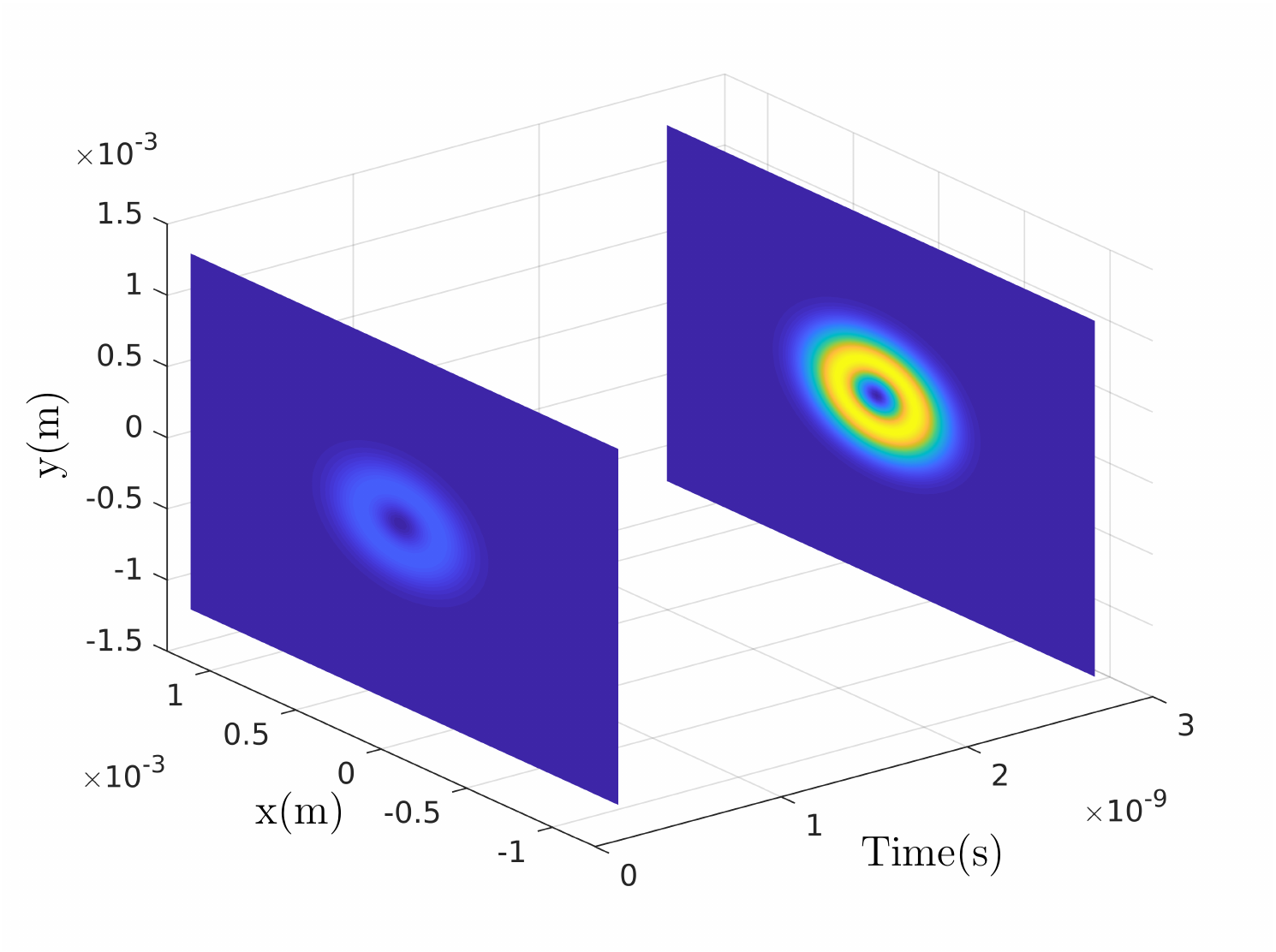}}
   \subfigure[\label{idlc}]{\includegraphics[scale=0.35]{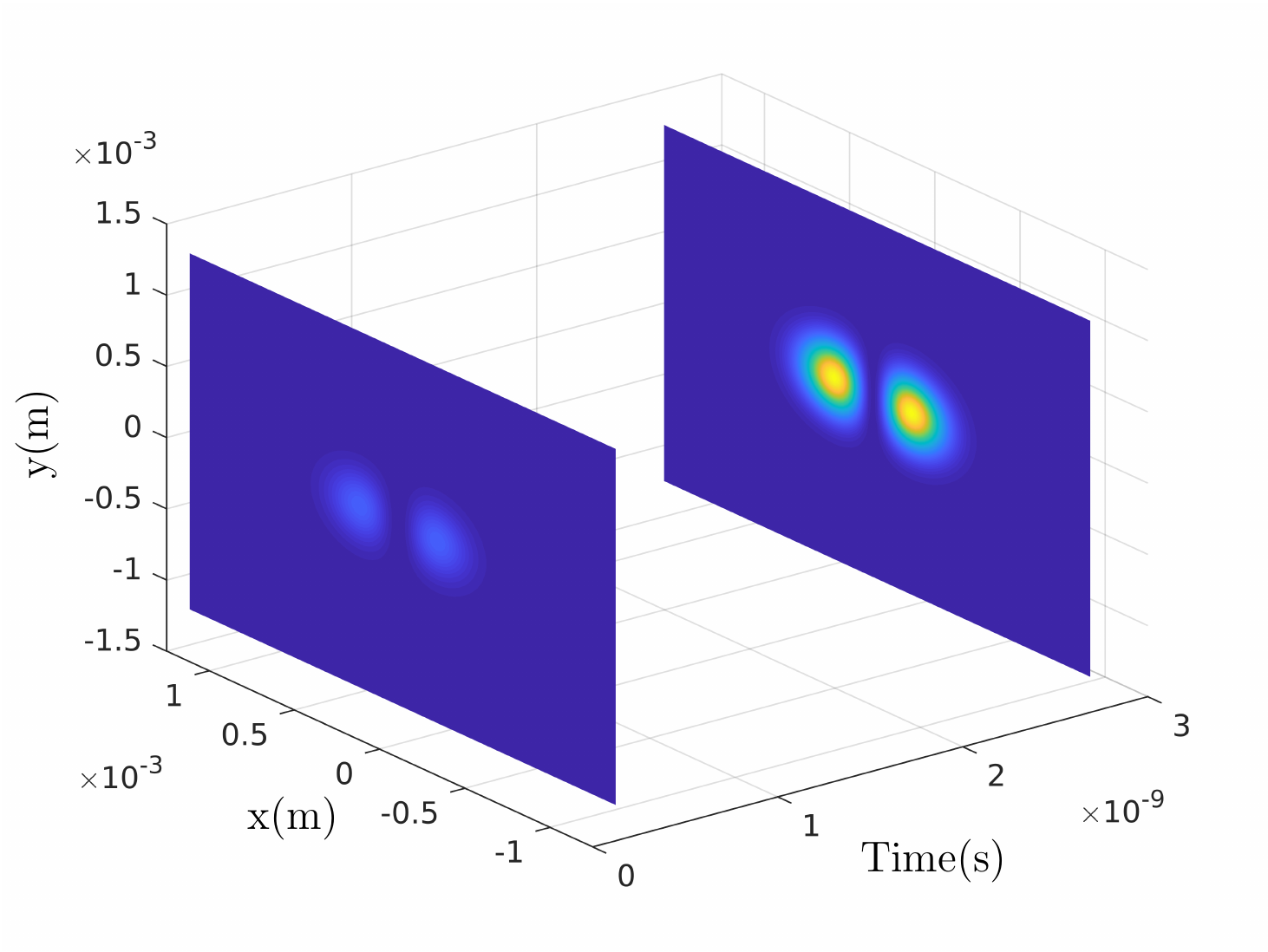}}
   \subfigure[\label{idli}]{\includegraphics[scale=0.35]{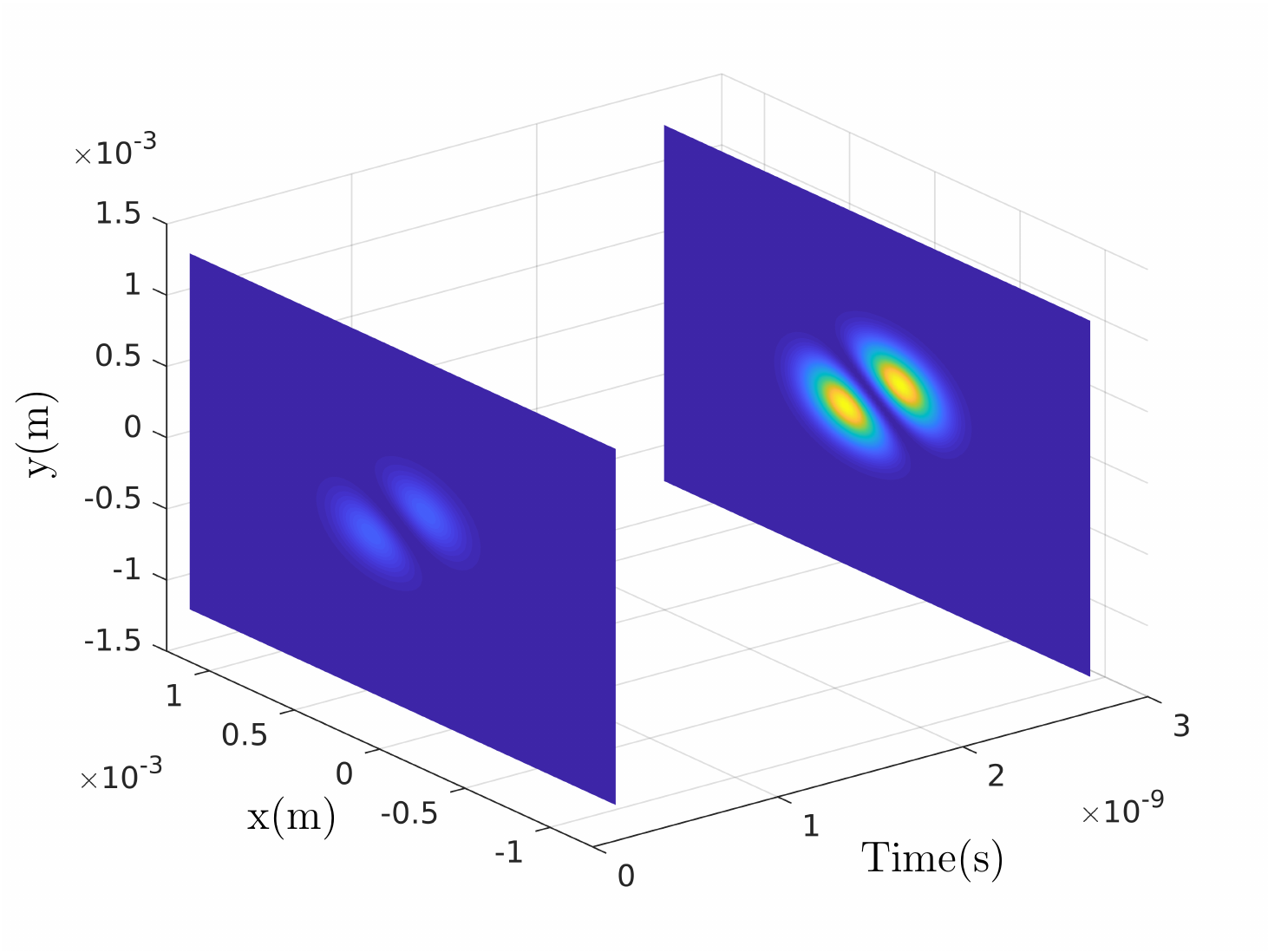}}
   \caption{Storing OAM in I-AFC. \subref{idl1}, \subref{idlc} and \subref{idli} Photon-echos for the ideal comb for the OAM states $\ket{1}$, $(\ket{1}+\ket{-1})/\sqrt{2}$ and $(\ket{1}+\mi \ket{-1})/\sqrt{2}$ respectively. The first and the second transverse profile in the plots correspond to the probability of the photon being unabsorbed and the probability of the photon-echo. All the calculations are done at $0$K.}\label{oam}
   \end{figure*}



Light carrying non-zero OAM interact differently with atomic ensemble. It has been shown that the OAM modes of light can impart torque on the atoms which can be used for optical tweezers.~\cite{allen92,andersen2006,padgett2011}. Furthermore, it has been shown that a moving atom interacting with a light beam carrying $\ell \hbar$ OAM experiences  an $\ell$ dependent Doppler shift in the azimuthal direction apart from the usual longitudinal Doppler shift $k v_z$. This azimuthal Doppler shift is entirely due to the OAM modes of light. The total Doppler shift for the OAM state $\ket{\ell}$  with an arbitrary polarization is given by \cite{allen94,allen95}
\begin{align}
\begin{aligned}
\delta_{LG}=&v_z\qty(-k+\dfrac{k r^2(z^2-z_R^2)}{2(z^2+z_R^2)^2}-\dfrac{(\abs{\ell}+1)z_R}{(z^2+z_R^2)})  \\
&-v_r\qty(\dfrac{kr}{\bar{z}})-v_\phi\qty(\dfrac{\ell}{r}) \label{ds}\\
=&\delta_{z}+\delta_{r}+\delta_{\phi},
\end{aligned}
\end{align}
where $v_r$, $v_\phi$ and $v_z$ are the radial, azimuthal and longitudinal component of the velocity $\vb{v}$ of the atom. 

Here, the term $k v_z$ is the usual Doppler shift along the propagation direction and will be the only term for a plane wave. The term $\propto (\abs{\ell}+1)$ is due to the Gouy phase of the LG mode and the terms including $v_z$ and $v_r$ are due to the transverse profile of the LG modes along the  radial direction~\cite{allen94}. The final $\ell$ dependent term which is directly proportional to the OAM of the LG mode accounts for the azimuthal Doppler shift. Typically, the $k v_z$ is the leading term dominating by a factor  $10^4$~\cite{prl2006} compared to the radial and azimuthal Doppler shifts. The above $\ell$-dependent Doppler shift yield an $\ell$ dependent phase in the paraxial light which might affect the fidelity of the OAM modes upon storage.  In Sec.~\ref{templg} we include the above Doppler shift $\delta_{LG}$ to incorporate the effect of temperature on the storage of OAM modes.

\section{Results}\label{Sec:Results}

In this section, we show that the vector-vortex states of light can be stored efficiently in an appropriately designed I-AFC based quantum memory. Here we start with storing the LG modes of a paraxial light in I-AFC. We show that if the number density $\mathcal{N}$ of the atomic ensemble is homogeneous then the LG modes can be stored  perfectly at low temperatures. At high temperatures the Doppler shift may affect the quality of storage.
In order to store polarization states of light, we need to prepare an ensemble which contains two frequency combs, corresponding to two orthogonal polarizations. We show that if the two frequency combs are identical then the  storage of the polarization states is perfect. Non-identical combs might result in imperfect storage. The I-AFC system capable of storing both, LG modes as well as the polarization states, can be employed to store vector-vortex beams. As an example, we show that I-AFC in Cs and Rb atoms are capable of storing vector-vortex modes.

\subsection{Storing LG modes in I-AFC}\label{lgs1}

LG modes are the eigenmodes of the paraxial wave equation in free space. However, in an atomic ensemble they might get affected due to the presence of induced atomic polarization $\mathcal{P}$, especially if the medium is inhomogeneous. In this section, we show that  an atomic ensemble possessing I-AFC can be used to store LG modes of light. In order to do so, we will solve the propagation  of the LG modes through such atomic ensemble and show that we observe a photon-echo at time $2\pi/\Delta$ with high efficiency, which is a signature of the I-AFC based quantum memory. Further, we show that the fidelity between the input and output states of light is near perfect for idealized case. 

The Hamiltonian of the atomic ensemble interacting with classical electromagnetic field can be written as 
\begin{equation}
\begin{split}
H=&\sum_{n=1}^{N_e}\hbar \omega_n^e\dyad{e_n}+\sum_{m=1}^{N_g}\hbar \omega_m^g\dyad{g_m}\\
&-\hbar \sum_{n,m}\qty(\Omega_{nm}\dyad{e_n}{g_m}e^{-\mi \omega_Lt}+H.C.),\label{hmlg}
\end{split}
\end{equation}
where $\hbar\omega_n^e$ is the energy of the $n$-th state in the excited level, $\hbar\omega_m^g$ is the energy of the $m$-state in the ground level, and $\Omega_{nm}=\dfrac{{d}_{nm}{{\mathcal{E}}}(\vb{r}_{\perp},z,t)}{2\hbar}$. Here  the electric field $\mathcal{E}(\vb{r}_{\perp},z,t)$ has the  mean frequency $\omega_L$.

The paraxial wave equation inside a medium can be written as~\cite{scully}
\begin{align}
  \qty[\laplacian_\perp+2\mi k\qty(\pdv{z}+\dfrac{1}{c}\pdv{t})]\mathcal{E}(\bm{r},t)&=-\dfrac{k^2}{\epsilon_0}\mathcal{P}(\bm{r},t),
\end{align}
where $\mathcal{P}$ is the induced atomic polarization. 
This equation can be solved formally for spatially homogeneous medium~(Appendix~\ref{lg}) and the expression for the output field reads
\begin{align}\label{l1}
  \begin{split}
{\mathcal{E}}&(\vb{r}_{\perp},z,t)=\int M(\vb{r_{\perp}}-\vb{r'_{\perp}},z)\times\\
&~\qty[\int N(t-\tau,z){\mathcal{E}}(\vb{r'_{\perp}},0,\tau-\dfrac{z}{c})d\tau]~d^2\vb{r'_{\perp}}.
\end{split}
\end{align}
where
\begin{align}
M(\vb{r_{\perp}},z)=&\dfrac{1}{2\pi}\int e^{\mi \vb{q}.\vb{r_{\perp}}}e^{-\mi q^2z/{2k}}d^2\vb{q},\\
N(t,z)=&\dfrac{1}{2\pi}\int e^{\mi \omega t}e^{-\mathcal{D}(\omega)z}dw.
\end{align}

If the input light is in a pure LG mode $LG_0^l(\vb{r_{\perp}})$ with temporal profile given by $\mathcal{E}(0,t)$, then the expression for the input electric field ${\mathcal{E}}(\vb{r}_{\perp},0,t)$ reads
\begin{align}
{\mathcal{E}}(\vb{r}_{\perp},0,t)= \mathcal{E}(0,t)LG_0^l(\vb{r_{\perp}}).\label{icv}
\end{align}
In this case,  the  output field ${\mathcal{E}}(\vb{r}_{\perp},z,t)$ will be 
\begin{equation}
\begin{split}
{\mathcal{E}}(\vb{r}_{\perp},z,t)&=\qty(\int M(\vb{r_{\perp}}-\vb{r'_{\perp}},z)  LG_0^l(\vb{r'_{\perp}}) ~d^2\vb{r'_{\perp}})\times\\
&\qty(\int N(t-\tau,z)~\mathcal{E}(0,\tau-z/c)d\tau).\label{l2}
\end{split}
\end{equation}
Interestingly, the evolution of the electric field  ${\mathcal{E}}(\vb{r}_{\perp},0,t)$ splits into two parts, one which drives the transverse evolution and other which drives the time evolution. On close inspection, we can see that the evolution in the transverse plane is identical to the one in vacuum~\eqref{alg}. Since, LG modes are the eigenmodes of paraxial wave-equation, the pure LG modes remain  unaffected in this evolution except acquiring an overall phase, i.e., Gouy phase given by given by $\exp[-\mi(2p+\abs{\ell}+1)\tan^{-1}(z/z_{R})]$ [Eq.~\ref{v1}]. Here  $z/z_{R} \sim 10^{-4}$ for the dimensions chosen for a typical quantum memory. Therefore, the Gouy phase become negligible which allows us to store a superposition of LG modes. 

Furthermore, the information about the I-AFC is completely contained in the kernel $N$ which controls the time evolution of the state. Therefore, the temporal part in Eq.~\eqref{l2} is identical to the ordinary I-AFC evolution which results in a photon-echo at times which are multiples of $2\pi/\Delta$, without affecting  the transverse part.  Therefore, I-AFC in the homogeneous atomic ensemble is fully capable of storing LG modes of light.

In Fig.~\ref{oam} we plot  the numerically obtained re-emission of LG modes from an ideal I-AFC. The comb spacing $\Delta$ here is $400$ MHz and peak width $\gamma$ is $5$ MHz. We choose $LG_0^1,~(LG_0^1+LG_0^{-1})/\sqrt{2}$ and $(LG_0^1+\mi LG_0^{-1})/\sqrt{2}$ modes with a Gaussian temporal profile. As expected, the LG modes rephase after time $2\pi/\Delta$ while preserving the transverse profile. The first transverse profile in Fig.~\ref{oam} at $\sim0.23$ns corresponds to the probability of the photon being unabsorbed, while the second transverse profile at $\sim2.7$ns represents the photon-echo corresponding to the input LG mode. The brightness of the transverse profile is proportional to the probability of the photon-emission which clearly indicates higher probability of the first-echo relative to the noise at $\sim0.23$ns. The optimized efficiency and fidelity are  obtained to be $53.44\%$ and $100\%$.

\begin{figure}
\includegraphics[scale=0.35]{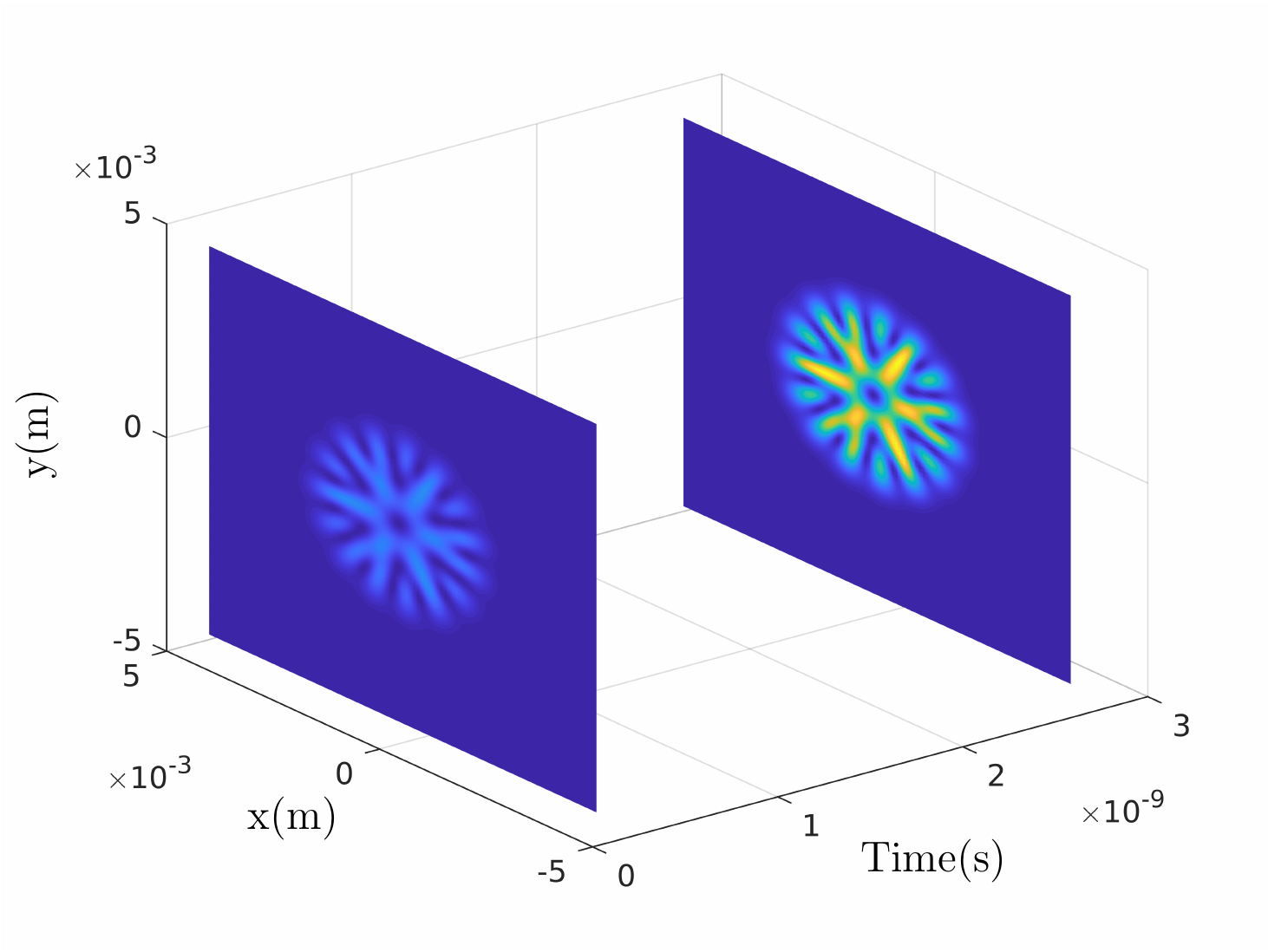}
\caption{Photon-echo for the ideal comb for the OAM state $(\ket{1}+\ket{-5}+\ket{10})/\sqrt{3}$.}
\label{hoam}
\end{figure}

Since for the parameters we have considered the Gouy phase is very small, one can store arbitrary superposition of higher dimensional LG modes  (within reasonable range of $\ell$ values) without affecting the fidelity. In Fig.~\ref{hoam} we show rephasing of $(LG_0^1+LG_0^{-5}+LG_0^{10})/\sqrt{3}$ state with $\sim 100 \%$ fidelity.

Note that, the storage of the LG modes was made possible by the assumption that the atomic number density $\mathcal{N}(\vb{r}_\perp)$ is homogeneous in the transverse plane, which made the kernel $N$ independent of transverse coordinates $\vb{r}_\perp$. Inhomogeneity in the atomic ensemble will affect the LG modes and the storage fidelity will not be perfect. We will discuss this in detail in Sec.~\ref{effect}.



\subsection{Quantum memory for polarization qubit}\label{polz}

In order to store polarization states of light, the system of interest must be capable of interacting with two orthogonal states of light identically. Here, we propose a scheme to store polarization states using I-AFC based quantum memory. For that purpose, we consider atoms with degenerate ground and excited states. The transitions between the ground states and excited states  satisfy the selection rules $\Delta m = \pm 1$. The Hamiltonian for such system interacting with light pulse of mean frequency $\omega_L$ reads
\begin{equation}
\begin{split}
H=&\sum_{n=1}^{N_e}\hbar \omega_n^e\dyad{e_n}+\sum_{m=1}^{N_g}\hbar \omega_m^g\dyad{g_m}\\
&-\hbar \sum_{n,m}\qty(\Omega_{nm}\dyad{e_n}{g_m}e^{-\mi \omega_Lt}+H.C.),\label{hm}
\end{split}
\end{equation}
where 
\begin{align}
\Omega_{nm}=\dfrac{\vb{d}_{nm}\vdot{\boldsymbol{\mathcal{E}}}(z,t)}{2\hbar},
\end{align}
$\vb{d}_{nm}= {d}_{nm}^+ \vu{e}_+ + {d}_{nm}^-\vu{e}_- \equiv\mqty[{ {d}_{nm}^+ \\  {d}_{nm}^-}]$ is the transition dipole moment vector between the $n$-th excited state $\ket{e_n}$ and $m$-th ground state $\ket{g_m}$ where the elements of the vector correspond to $\Delta m = \pm 1$ transitions.

The electric field vector  $\bm{\mathcal{E}}(z,t)$ in a superposition of the two polarization can be written as \cite{bransden2003}
\begin{align}
{\boldsymbol{\mathcal{E}}(z,t)}&={\mathcal{E}}_+(z,t)\vu{e}_++{\mathcal{E}}_-(z,t)\vu{e}_-\equiv\mqty[{{\mathcal{E}}_+ \\ {\mathcal{E}}_-}],\label{e2}
\end{align}
where the $\vu{e}_{\pm}$  are the unit vectors along left and right circular polarization and interact with the transition corresponding to $\Delta m=\pm1$. 

The dynamics of the electric field and the atomic ensemble is given by Maxwell-SchrÃ¶dinger equations [Eqs.~\eqref{e11} and \eqref{e22}]. Solving the dynamics for the electric field vectors results in expression for the output electric field $\tilde{\bm{\mathcal E}}$ in frequency domain,
\begin{align}
  \pdv{\tilde{\bm{\mathcal E}}}{z}=-\dfrac{\mi\omega}{c}\tilde{\bm{\mathcal{E}}}-\dfrac{\omega_L\mathcal{N}}{2 c\hbar\epsilon_0}\sum_{n,m}\dfrac{\vb{d}_{nm}(\vb{d}_{nm}\vdot{\tilde{\bm{\mathcal{E}})}}}{\qty[\mi(\Delta_{nm}+\omega)+\dfrac{\gamma}{2}]}{\rho}_{mm},\label{6}
\end{align}
where $\rho_{mm}$ is the population of the $m$-th energy level. On using Eq.~(\ref{e2}) and (\ref{6}), the equations for the two orthogonal polarization components can be written as
\begin{align}
\pdv{z}  \mqty[{\tilde{\mathcal{E}}_+ \\ \tilde{\mathcal{E}}_-}] &=\mqty[{-\mi \omega}/{c}-\mathcal{D}^+(\omega) & -\mathcal{G}(\omega) \\ 
-\mathcal{G}(\omega) & {-\mi \omega}/{c}-\mathcal{D}^-(\omega)
] \mqty[{\tilde{\mathcal{E}}_+ \\ \tilde{\mathcal{E}}_-}],\label{b1}\\
  &   \equiv A  \mqty[{\tilde{\mathcal{E}}_+ \\ \tilde{\mathcal{E}}_-}].
\end{align}

\noindent From Eq.~\eqref{b1} we can calculate the output electric field which reads as $\tilde{\bm{\mathcal{E}}}(z,\omega) = \text{e}^{Az}\tilde{\bm{\mathcal{E}}}(0,\omega) $. Here, $\mathcal{D}^\pm(\omega)$ and $\mathcal{G}(\omega)$ are defined as
\begin{align}
\mathcal{D}^\pm(\omega)&=\sum_{nm}{\dfrac{g_{mm}}{\qty[\mi(\Delta_{nm}+\omega)+\dfrac{\gamma}{2}]}{d^{\pm 2}_{nm}}}, \label{tp}\\
  \mathcal{G}(\omega)&=\sum_{nm}{\dfrac{g_{mm}}{\qty[\mi(\Delta_{nm}+\omega)+\dfrac{\gamma}{2}]}d_{nm}^{+}d^-_{nm}}.
\end{align}

For the case, when the magnetic quantum number $m$ is not a good quantum number for atomic states, both the transition dipole moment $d^\pm_{nm}$ between $n$-th excited state and the $m$-th ground state might not vanish. This will result in non-zero $\mathcal{G}(\omega)$ term which is responsible for the mixing of the two polarizations. On the other hand, the terms $\mathcal{D}^\pm(\omega)$ in Eq.~(\ref{b1}) are the propagators corresponding to the two I-AFCs corresponding to $\Delta m = \pm 1$ transitions. For the case when the off-diagonal term vanishes, the two orthogonal polarizations propagate independently through the I-AFC.

Note that if $D^{\pm}(\omega)$ is same for both polarizations and $\mathcal{G} = 0$, then the matrix $A$ is proportional to identity matrix. Hence, the propagation of the light inside the I-AFC will be independent of the polarization, resulting in polarization independent storage. In such cases, the I-AFC based quantum memory can store the polarization efficiently. However, in physical systems the propagators for the two combs may not always be equal. Different propagators $\mathcal{D}^\pm$ may result in different photon-echo times and different efficiencies  for orthogonal polarizations, which in turn may result in lower fidelity between the input and output states of light if the input light is in some superposition of the two polarizations.

In conclusion,  an atomic ensemble containing two identical I-AFC corresponding to two orthogonal polarizations can store polarization states of light efficiently. Since, the OAM states of light (LG modes) are independent of polarization, and if the atomic ensemble is homogeneous, it can be used to store these states also efficiently. Therefore, one  can store vector-vortex states of light in I-AFC based quantum memory. 

\subsection{Factors affecting the quality of quantum memory}\label{effect}
So far we have discussed the storage of vector-vortex modes only in ideal systems. There are several factors which might affect the quality of the storage. In this section, we discuss a few of those factors and their effects in detail. 


\begin{figure}[]
\subfigure[\label{nvari}]{\includegraphics[height=3.7cm,width=4.2cm]{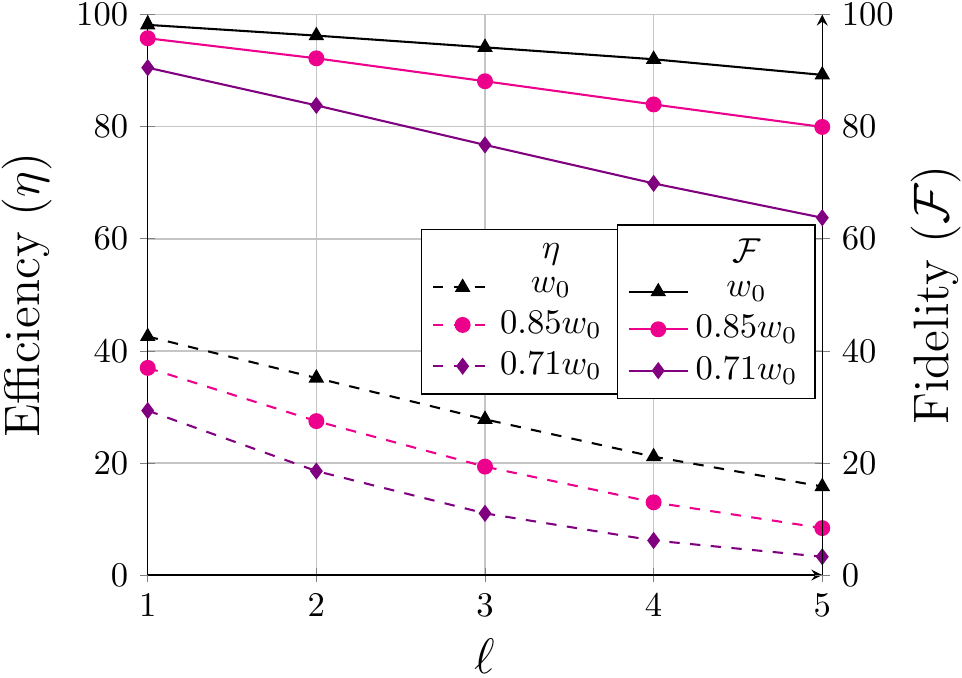}}
\subfigure[\label{tempvi}]{\includegraphics[height=3.7cm,width=4.2cm]{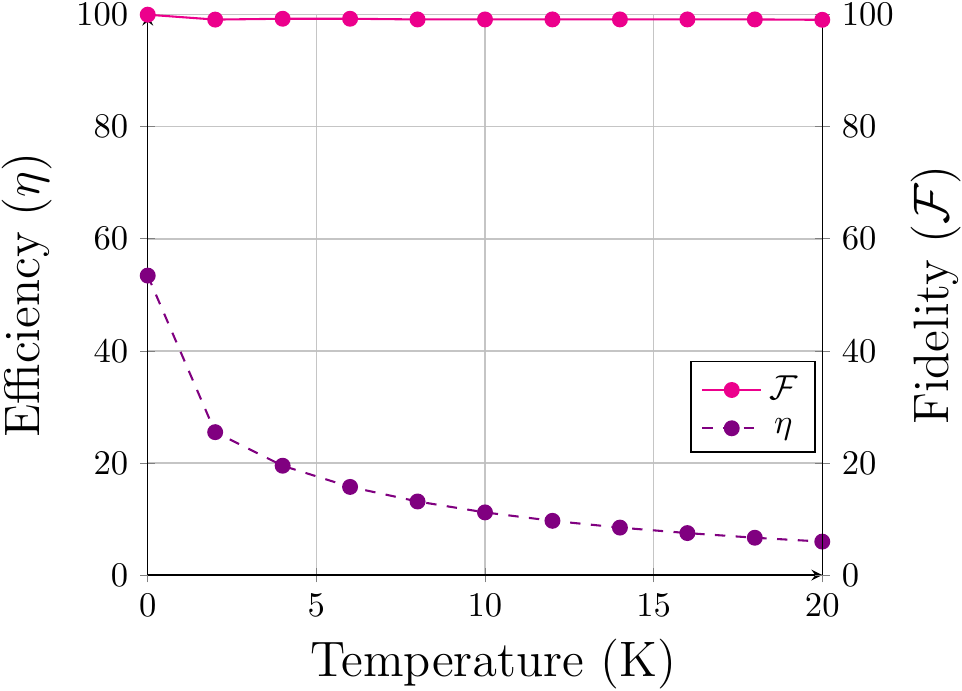}}
\caption{ Here we plot the efficiency (dashed curve) and the fidelity (solid curve) for different values of $w_0'$ as a function of $\ell$ for the input state $\ket{\ell}$ in \subref{nvari} and  as a function of temperature for the OAM state $(\ket{1}+\ket{-1})/\sqrt{2}$ in \subref{tempvi}. For these plots,  we consider an ideal comb consisting of nine peaks with uniform comb spacing $\Delta=400$MHz and peak width $\gamma=5$MHz.}
\label{idloam}
 \end{figure} 

\subsubsection{Effects of non-homogeneous number density on the storage of  LG modes}\label{nvlg}

As we have noticed in the case of storing LG modes in homogeneous systems, the evolution of the electric field decouples in two parts [Eq.~\eqref{l2}]; one corresponds to the evolution in the transverse plane and other corresponding to the time evolution. 
However,  if the atomic number density is not homogeneous this separation is not guaranteed. To study the adverse effect of non-homogeneous number density on the quality of the quantum memory, we consider a simple case where the number density is a function of $|\vb{r}_\perp| \equiv r_{\perp} = \sqrt{x^2+y^2}$, in the transverse plane. For simplicity, we choose a Gaussian distribution of atomic density in the transverse plane, i.e., $\mathcal{N'}(\vb{r}_{\perp})=\mathcal{N}_0 \exp[-{(x^2+y^2)}/{2 w_0'^2}]$, where $w_0'$ is the width of the distribution and $\mathcal{N}_0$ is a constant.

In order to calculate the effect of non-homogeneous number density, we need to solve  Eq.~\eqref{Eq:ETilde}, which can be done numerically. The efficiency of the photon-echo and the fidelity between the input and output states of the electric field can be calculated using the relations given in Eq.~\eqref{eff} and Eq.~\eqref{fidp}.  In fig.~\ref{nvari} we plot the efficiency $\eta$ and the fidelity $\mathcal{F}$ as a function of the $\ell$ value of the LG modes for different values of $w_0'$ for a fixed beam waist $w_0$, while keeping $p=0$.

 As expected, the non-homogeneity of atomic ensemble affects the quality of the quantum memory more for the larger values of $\ell$. This result can be explained by noticing that the  size of the tranverse profile of the LG modes increases with increasing the $\ell$ value. Hence, the beams with large $\ell$ value  see fewer atoms which results in a lower absorption rate and hence lower efficiency. Furthermore, the LG modes are not the eigenmodes inside the non-homogeneous medium. This explains the drop in the fidelity.


\subsubsection{Effects of temperature  on the storage of  LG modes}\label{templg}

Besides the non-homogeneous number density, thermal effects can also affect the quality of the storage of OAM states of light. To study the effect of the temperature on the OAM storage, we consider a homogeneous atomic ensemble in a thermal equilibrium at temperature $T$.
The atomic velocity distribution in such ensemble can be written as
\begin{align}
p(v)d^3v=\qty(\dfrac{m}{2\pi k_bT})^{3/2}\exp(-m(v_x^2+v_y^2+v_z^2)/2k_BT)d^3v,\label{pdt}
\end{align}
where $m$ is the mass of the atom, $v_x,v_y$ and $v_z$ are the velocity of atom along the $x$, $y$, and $z$ directions, respectively, and $k_B$ is the Boltzmann constant.

As we have discussed in Sec.~\ref{dp}, an atom moving with velocity $\vb{v}$ and interacting with light with non-zero OAM value experiences a change in the detuning from $\Delta$ to $\Delta+\delta_{LG}(\vb{v})$ due to the Doppler shift. Hence the modified expression for the atomic polarization $\tilde{\mathcal{P}}(\vb{r}_{\perp},z,\omega,\vb{v})$ after incorporating  the Doppler shift can be written as
\begin{align}
\tilde{\mathcal{P}}(\vb{r}_{\perp},z,\omega,\vb{v})=2{\mathcal{N}}\sum_{n,m}\dfrac{\mi d_{nm}^2\tilde{\mathcal{E}}(\vb{q},z,\omega){\rho}_{mm}}{2\hbar\qty[\mi(\Delta_{nm}+\delta_{LG}(\vb{v})+\omega)+\dfrac{\gamma}{2}]}.\label{p1}
\end{align}
The net atomic polarization can be calculated by averaging $\tilde{\mathcal{P}}(\vb{r}_{\perp},z,\omega,\vb{v})$ over all the velocities~\cite{mandel1995,iafc19}, i.e., 
 \begin{align}
\tilde{\mathcal{P}}(\vb{r}_{\perp},z,\omega)=&\int\tilde{\mathcal{P}}(\vb{r}_{\perp},z,\omega,v)p_vd^3v.\label{avg}
\end{align}
To calculate the analytical expression from Eq.~\eqref{avg}, the Gaussian distribution $p_v \propto\exp[-{m v_i^2}/{2 k_B T}]$ is approximated by the corresponding Lorentzian distribution $\mathcal{L}_v=\dfrac{a}{\pi (a^2+v_i^2)}$ where  $a\propto\sqrt{\dfrac{k_B T}{m}}$~\citep{iafc19}. The proportionality constant for `$a$' is obtained using numerical curve fitting and found to be close to $0.76$. The final expression for the average macroscopic polarization can be written as
\begin{align}
\tilde{\mathcal{P}}(\vb{r}_{\perp},z,\omega)
 =&2{\mathcal{N}}\sum_{n,m}\dfrac{\mi d_{nm}^{2}{\tilde{{\mathcal{E}}}(\vb{q},z,\omega)}{\rho}_{mm}}{2\hbar\qty[\mi(\Delta_{nm}+\omega)+\dfrac{\gamma}{2}+a f(\vb{r}_{\perp},z)]},\label{p2}
\end{align}
where the function $f(\vb{r}_{\perp},z)$ reads
  \begin{align}
    \begin{split}
f(\vb{r}_{\perp},z)=&\dfrac{k(x+y)}{\bar z}+\dfrac{\ell (x-y)}{x^2+y^2}
-\dfrac{(2p+\abs{\ell}+1)z_R}{z^2+z_R^2}\\& -\dfrac{k(x^2+y^2)}{2\bar z^2}\dfrac{(z^2-z_R^2)}{z^2}+k.
\end{split}\label{p3}
\end{align}

From Eq.~\eqref{p2}, it is clear that the effect of the temperature shows up as the broadening of the peak width $\gamma$ by a factor $af(\vb{r}_\perp,z)$ where $ka$ is the leading order term. This leading term results in lowering the finesse of the frequency comb and result in lower efficiency of the quantum memory~\citep{iafc19}. Other sub-leading terms in the broadening are $\ell$-dependent  and position-dependent due to the transverse profile of the field. The $\ell$- and position-dependence of these terms will yield different efficiencies for different LG-modes which will affect the fidelity of the quantum memory.

To study the effect of the Doppler shift $\delta_{LG}$ due to the transverse profile of light, we numerically solve the dynamical equations for the atomic coherence \eqref{pe} and the electric field \eqref{dy} by replacing $\qty(i\Delta_{nm}+\dfrac{\gamma}{2})$ with  $\qty(i\Delta_{nm}+\dfrac{\gamma}{2}+a f(\vb{r}_{\perp},z))$ in \eqref{dy}.
The variation of the optimized efficiency and the corresponding fidelity  with the temperature for the ideal comb is shown in Fig.~\ref{tempvi}. 
In this figure, we consider an ideal comb with $9$ teeth and the tooth spacing $\Delta=400$MHz. The peak width $\gamma = 5$MHz. We see that the efficiency of the quantum memory drops as we increase the temperature falling below $10\%$ for $T=20$K. However, the fidelity appears to be unaffected. This shows that the contribution from the $ka$ term is much stronger than the contribution from the sub-leading terms in the Doppler shift.


\subsubsection{Factors affecting polarization storage}
Unlike the LG modes, polarization states of light do not depend on the spatial coordinates as long as the transverse plane is well defined. Therefore, the non-homogeneity in  the number density has very little effect on polarization storage. However, factors such as unequal $\mathcal{D}$ propagators, and the  non-overlapping frequency combs corresponding to two orthogonal polarizations can affect the quality of the quantum memory. Here we will discuss a few of those factors and study their effects. For simplicity, we will assume $\mathcal{G} = 0$ throughout this section as most of the real systems exhibit this property.

\begin{figure*}\label{pol}
\subfigure[\label{figa}]{\includegraphics[scale=0.45]{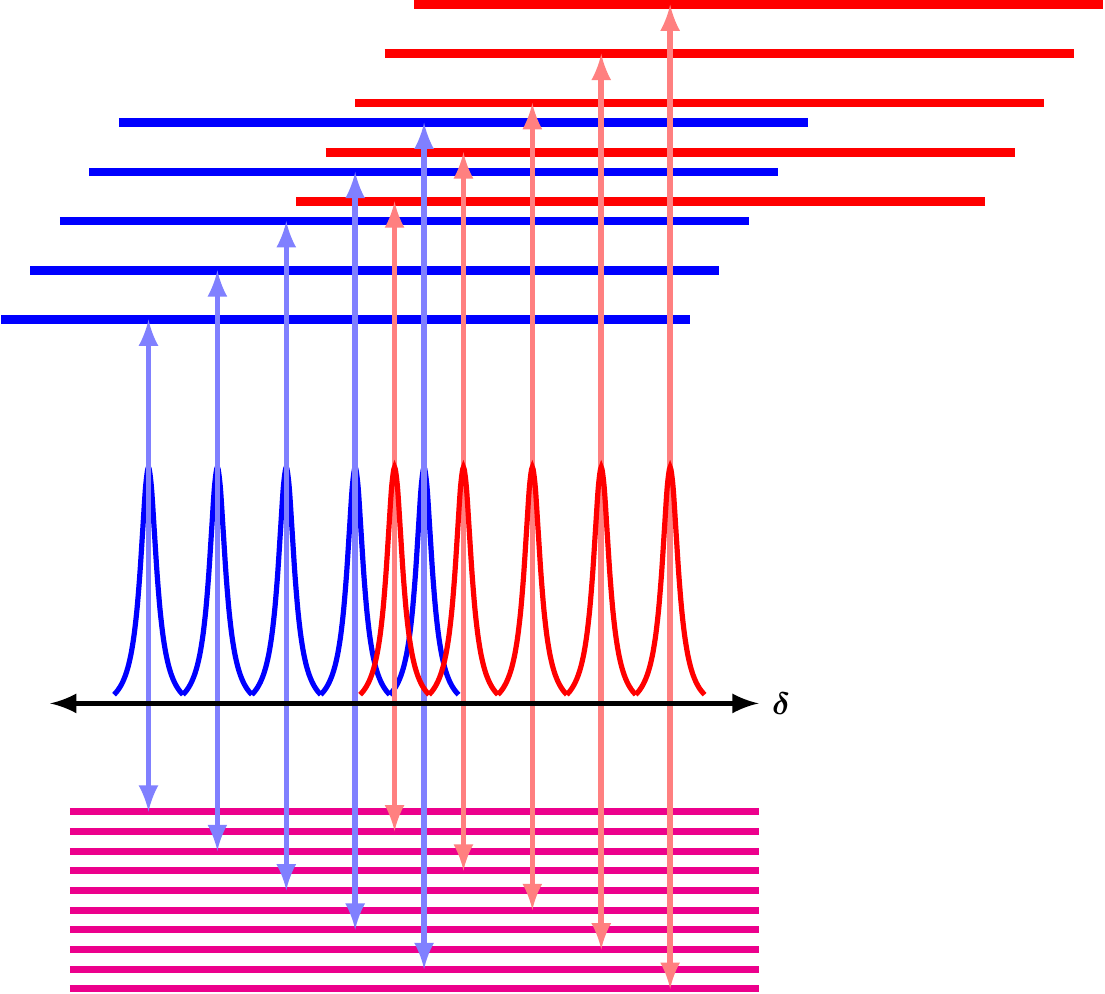}}
\subfigure[\label{fig2}]{\includegraphics[scale=0.62]{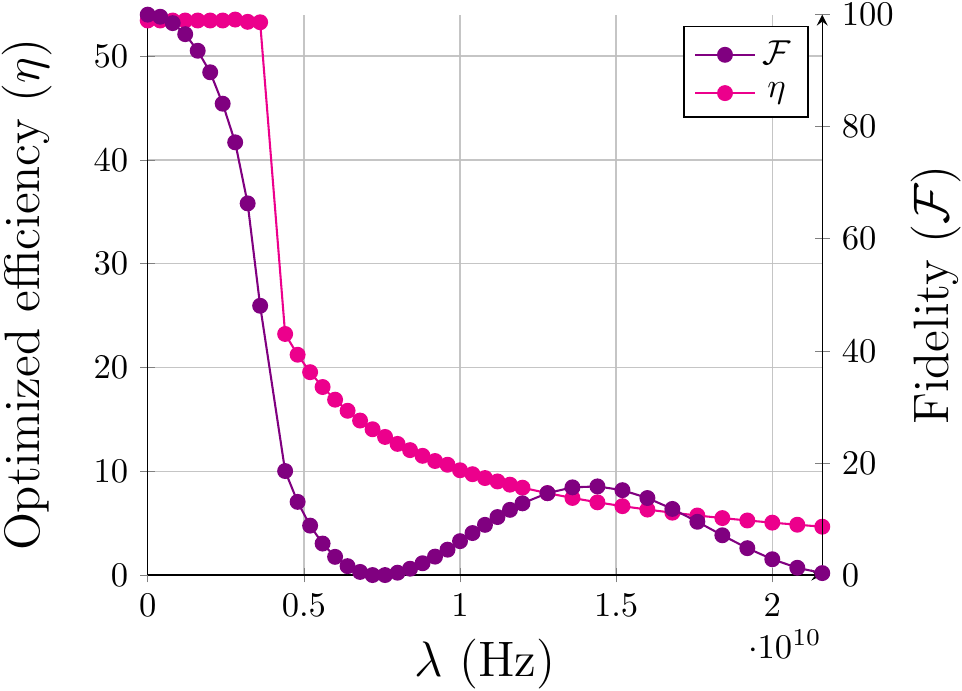}}
\subfigure[\label{fig2a}]{\includegraphics[scale=0.62]{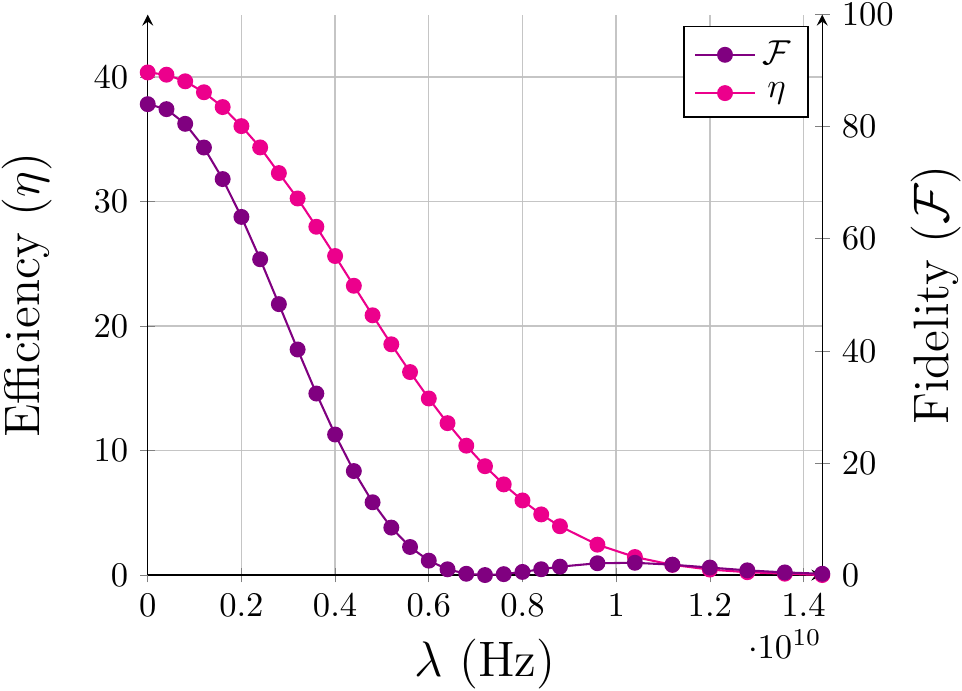}}
\caption{Storing polarization in I-AFC. \subref{figa} Storing plarization states using I-AFC. The two orthogonal polarization components couple separately to two non-overlapping frequency combs (red and blue) according to the transition selection rule $\Delta m=\pm1$. \subref{fig2} Variation of the optimized efficiency ($\eta$) w.r.t the input width and the corresponding fidelity $(\mathcal{F})$. \subref{fig2a} Variation of the storage efficiency ($\eta$) and the corresponding fidelity $(\mathcal{F})$  as a function of separation between two combs at a fixed input width of $2.4$ GHz. Both frequency combs consist of 11 peaks with fixed comb spacing of 400 MHz and peak width 5 MHz. Here the input field is taken as $\mathcal{E}(0,t)=e^{-\omega^2/(2 b^2)}$.}
\end{figure*}

The most common factor that can affect the quality of the quantum memory for polarization states is the unequal $\mathcal{D}$ propagators, i.e., $\mathcal{D}^+ \ne \mathcal{D}^-$. This will result in different echo times and efficiencies for the two orthogonal polarizations. 
By choosing a wave-plate appropriately, one can compensate for the unequal storage time for the two polarizations. However, the different efficiencies of the two polarization components can affect the fidelity of the output  polarization states. This can be compensated  by choosing the mean frequency of light $\omega_L$ lying exactly in the middle of the two combs, or by selectively absorbing light corresponding to a particular polarization which will result in the overall lower efficiency but higher fidelity.

In some cases, even if the frequency combs corresponding to two polarizations are identical, they might be displaced with respect to each other (Fig.~\ref{figa}). This factor can also adversely affect the storage of polarization states. In such cases, the photon-echo for both the polarizations occur at the same time but an additional phase is attributed to them due to the shifted combs. This can be understood as follows: 
Consider a comb shifted by  $\pm\lambda/2$  so that the detuning  $\Delta_{nm}$ becomes  $\rightarrow  \Delta_{nm}\pm \lambda/2$ for two combs. Solving equations  \eqref{e11} and \eqref{e22} in the frequency domain yield
 \begin{align}
\tilde{{\mathcal{E}}}_\pm(z,\omega_\pm)=&e^{-\mathcal{D^\pm} z} e^{-\mi\omega z/c}e^{\mp \mi\lambda z/2c} \tilde{{\mathcal{E}}}(0,\omega_\pm),
\end{align}
where $\tilde{{\mathcal{E}}}(0,\omega_\pm)$ is the input electric field amplitude, $\omega_\pm = \omega \pm \lambda/2$, and $\mathcal{D}$ is given by
\begin{align}
\mathcal{D}^\pm(\omega_\pm)=\sum_{n,m}\dfrac{g_{mm}}{\qty[\mi(\Delta_{nm}+\omega_\pm)+\dfrac{\gamma}{2}]}d_{nm}^2,
\end{align}
Hence, the output fields get equal and oppposite phase $e^{\mp\mi \lambda z/2c}$ for two combs. For small shift $\lambda$, the fidelity drop is also small whereas, the drop increases as the value of $\lambda$ increases.

We consider an I-AFC for polarization storage such that the frequency combs corresponding to two orthogonal polarizations are ideal but displaced with respect to each other by a magnitude $\lambda$. In Fig.~\ref{fig2} and \ref{fig2a}, we plot the effect of the relative shift $\lambda$  on the  efficiency of the quantum memory and the fidelity of the output polarization state with respect to the input state.  In these results, each of the frequency comb has eleven teeth with comb spacing $\Delta = 400$ MHz and peak width $\gamma = 5$MHz. Hence, the total size of each of the comb is $4$GHz. These parameters are close to  $6s_{1/2} \leftrightarrow 8p_{3/2}$ transition in Cs atom.

Here we consider two cases to study the effect of $\lambda$. In the first case, the intensity is optimized over the spectral width and the mean frequency of the incoming light and the corresponding fidelity is obtained as shown in Fig.~\ref{fig2}. From this figure, we can see that efficiency $\eta$ is close to $54\%$ when the two combs are perfectly overlapping. This is the maximum efficiency that can be achieved in I-AFC for forward propagating modes. This feature persists as long as the separation between the two combs is $\lambda < 4$GHz. After that point the efficiency start to decrease. Interestingly, the drop in the efficiency is sharp. The value of $\eta$ drops from $\sim54\%$ at $\lambda = 3.6$GHz to below $24\%$ at $\lambda = 4.4$GHz.  On the other hand, the Fidelity $\mathcal{F}$ shows a smooth behaviour as we increase $\lambda$. It starts with fidelity $\mathcal{F} = 100\%$ at $\lambda = 0$ and show a damped oscillations as we increase $\lambda$.

In the second case, the optimization is done over the mean frequency of input keeping the spectral width of the input to be fixed [see Fig.~\ref{fig2a}].
In Fig.~\ref{fig2a} the input width is fixed at $2.5$ GHz.  Contrary to the previous case, here we can see a smooth variation in the efficiency as we increase $\lambda$. However, since the spectral width is fixed, the efficiency  $\eta\sim40\%$ at $\lambda = 0$~GHz which is considerably lower than the maximum possible, i.e., $54\%$. Similarly, the fidelity  $\mathcal{F}\sim 84\%$ at $\lambda = 0$.

Apart from these factors, temperature may also affect the storage efficiency and fidelity. However, those effects are generally independent of the polarization and affect only the overall efficiency, not the fidelity. In conclusion, the I-AFC based quantum memory for vector-vortex states of light is robust and efficient against prominent environmental factors.

\begin{figure}
\subfigure[\label{cs-tr}]{\includegraphics[width=4.2cm]{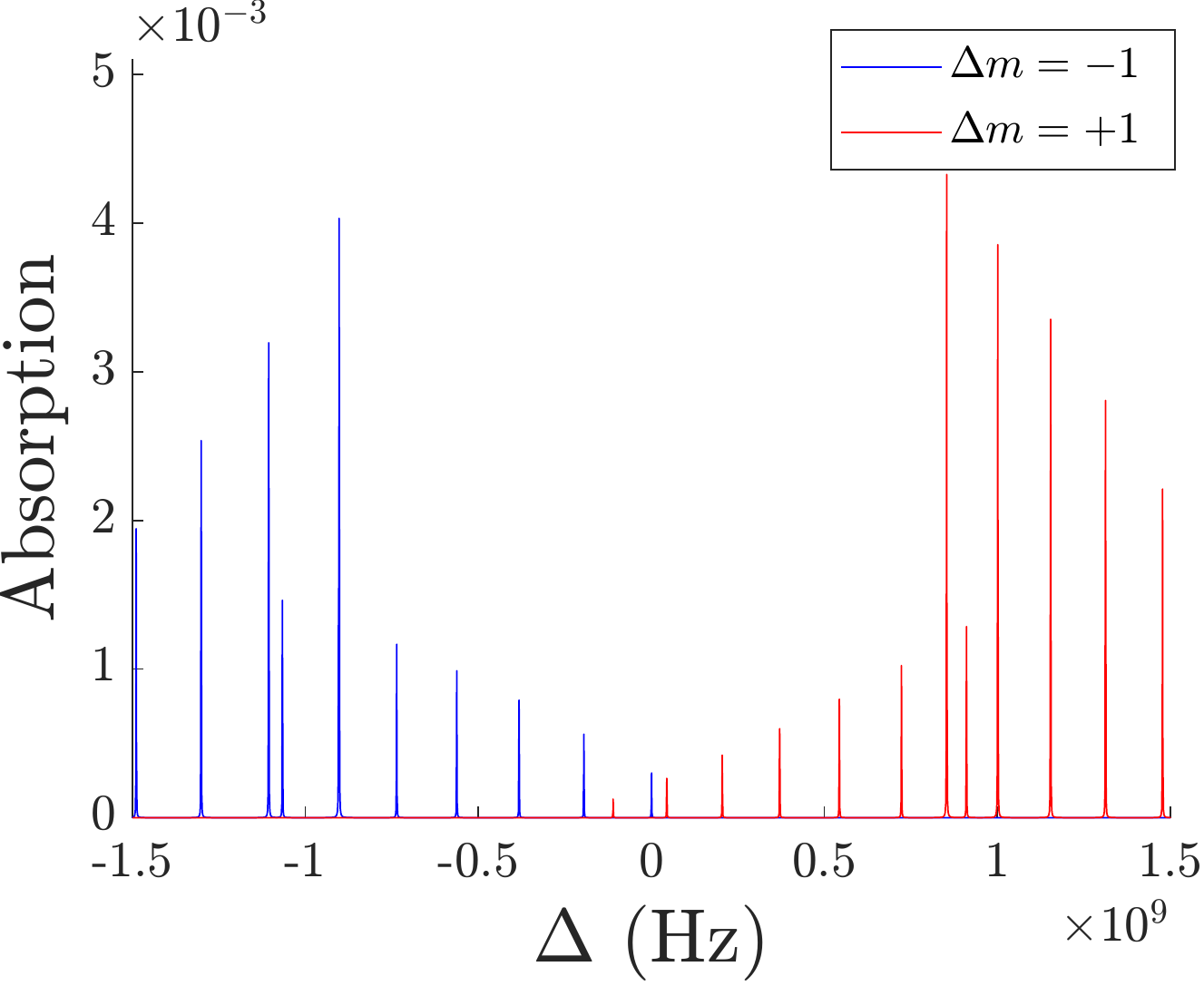}}
\subfigure[\label{rb-tr}]{\includegraphics[width=4.2cm]{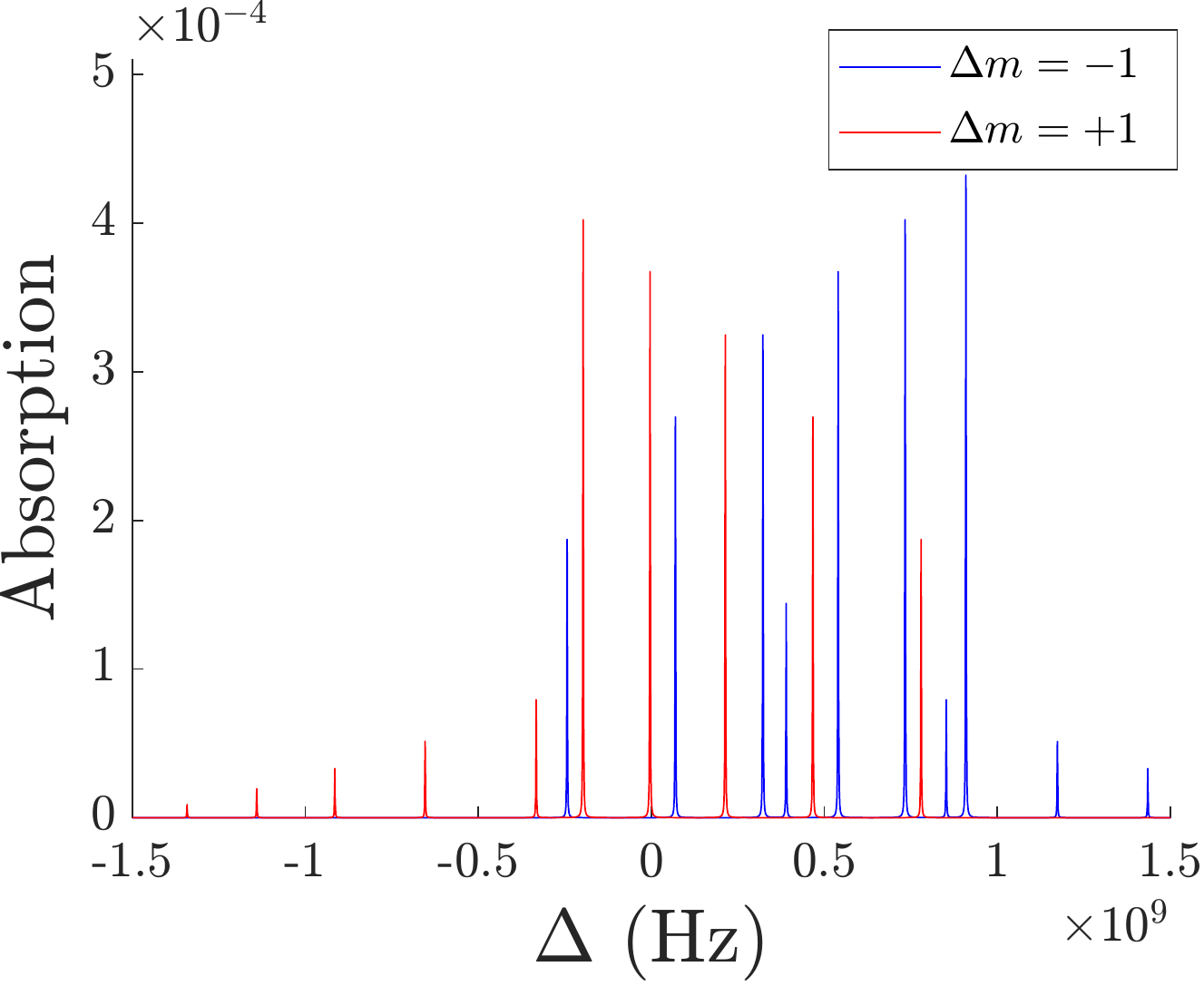}}  
\subfigure[\label{cspl}]{\includegraphics[width=4.2cm]{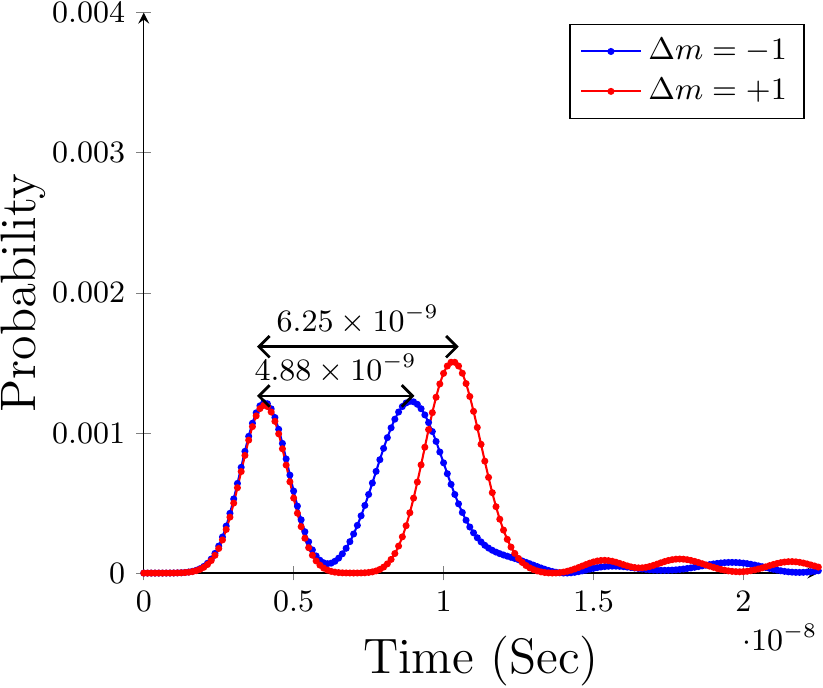}}
\subfigure[\label{rbpl}]{\includegraphics[width=4.2cm]{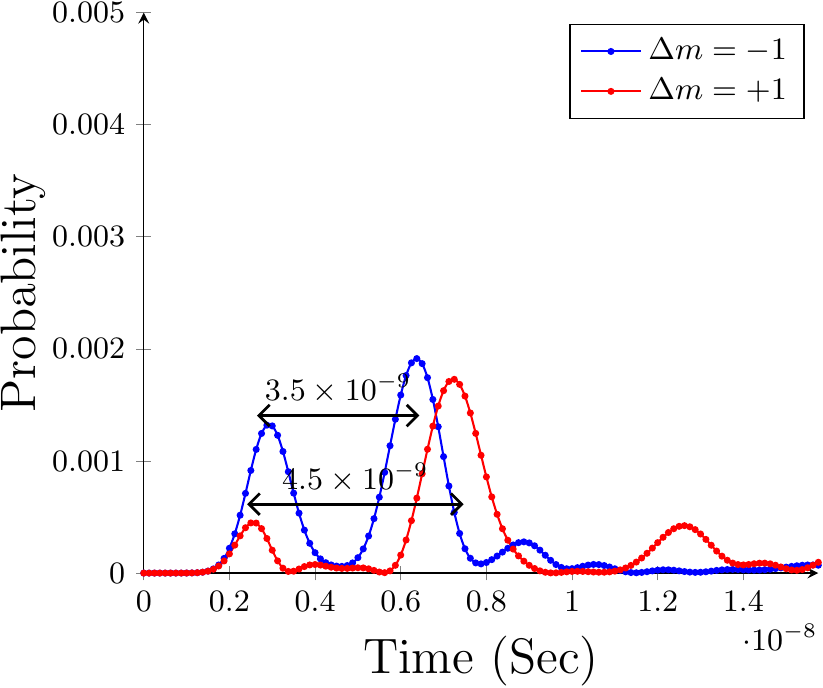}}

\caption{I-AFC in Cs and Rb atoms. In \subref{cs-tr} and \subref{rb-tr} we plot the frequency combs in the Cs and Rb atoms. Here $6s_{1/2}\leftrightarrow 8p_{3/2}$ and $5s_{1/2}\leftrightarrow 6p_{3/2}$ transitions are considered for Cs and Rb atoms, respectively. The degeneracy is lifted by applying external magnetic field of strength $0.05$T and $0.06$T, respectively, for the two cases. In \subref{cspl} and \subref{rbpl} we plot the photon-echo corresponding to $\Delta m=+1$ and $\Delta m=-1$ transitions in Cs and Rb atoms respectively.}
 \label{poam}
\end{figure}

\subsection{I-AFC in Cesium and Rubidium atoms}\label{csrb}
 
 
In this section we show that the Cesium and Rubidium atoms can be a feasible system  to store both the polarization as well as OAM modes of light. Hence, these atoms are suitable for storing vector-vortex states of light.

We consider the  $6s_{1/2}\leftrightarrow 8p_{3/2}$ transitions for Cs atom and  $5s_{1/2}\leftrightarrow 6p_{3/2}$ transition for Rb atom. In both the atoms, the atomic transitions are such that only one of the transition dipole moments $d_{nm}^+$ or $d_{nm}^-$ is non zero, thus giving $\mathcal{G}(\omega)=0$ which results in an independent propagation of the two polarization components. In both the atoms, the frequency combs are not uniform and the combs corresponding to $\Delta m = \pm 1$ are shifted with respect to each other (see Figs.~\ref{cs-tr} and~\ref{rb-tr}).

In Fig.~\ref{cspl} (Cs) and \ref{rbpl} (Rb), we show numerically obtained photon-echo for the polarization storage. The optimized fidelity and efficiency $(\mathcal{F},\eta)$ for the Cs and Rb atom are obtained to be  $(88.23\%,41.3\%)$ and $(87.46\%,41.16\%)$ respectively. Since, the $\mathcal{D}^\pm(\omega)$ propagators are non-identical for both Cs and Rb atoms, the efficiency and fidelity in these cases is lower than the ideal cases.
 Similarly, we observe the rephasing of the LG modes in Cs and Rb atoms. The optimized parameters, $(\mathcal{F},\eta)$ in this case are obtained to be $(98.7\%~51.96\%)$ for Cs and $(97.91\%~51.84\%)$ for the Rb atom and the photon-echo plots are similar to Fig.~\ref{oam}.

\begin{figure}[]
\subfigure[\label{nvar}]{\includegraphics[height=3.7cm,width=4.2cm]{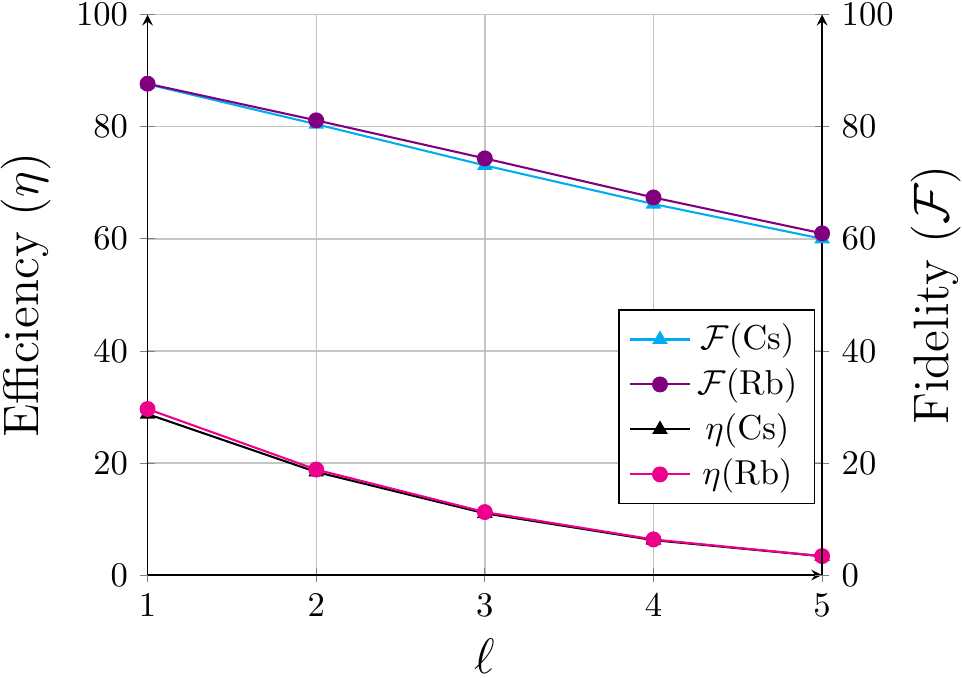}}
\subfigure[\label{tempv}]{\includegraphics[height=3.7cm,width=4.2cm]{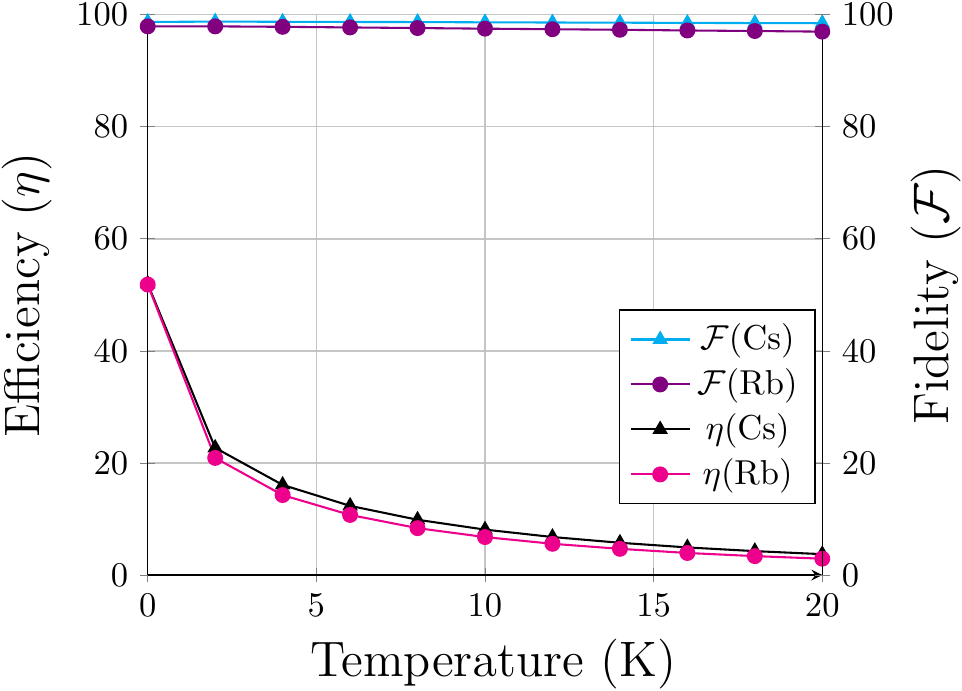}}
\caption{Effect of non-homogeneous number density and temperature on OAM storage. \subref{nvar} Efficiency and the corresponding fidelity as a function of $\ell$ value for Cs and Rb atoms for the input state $\ket{\ell}$.
 \subref{tempv} Efficiency and the corresponding fidelity in Cs and Rb atoms as a function of temperature for the OAM state $(\ket{1}+\ket{-1})/\sqrt{2}$.}\label{toam}
 \end{figure}

We also show the effects of non-homogeneous number density on the OAM storage in case of Cs and Rb atoms in Fig.~\ref{nvar} where {$\mathcal{N'}(\vb{r}_{\perp})=\mathcal{N} \exp[-{(x^2+y^2)}/{2 w_0'^2}]$} with $w_0'=0.71w_0$. It is clear that the Cs and Rb atoms show a similar drop for the efficiency and fidelity as in the ideal case discussed in Sec.~\ref{nvlg} [Fig.~{\ref{nvari}}]. Fig.~\ref{tempv} shows the effect of the temperature on storing OAM modes in Cs and Rb and the results are similar as that of an ideal comb (see Fig.~{\ref{tempvi}}).

Since Cesium and Rubidum atoms are capable of storing both, the polarization and OAM modes, we expect it to store the vector vortex beams. 

\section{Conclusion}\label{Sec:Conclusion}
Storing the internal states such as polarization and the OAM modes, and the vector-vortex states of light is essential for long range quantum communication and quantum information processing. In this paper, we propose the I-AFC based quantum memory to store the vector-vortex beams of light. We show that an atomic ensemble with I-AFC and homogeneous number density in the transverse plane is capable of storing the LG modes of light efficiently. Moreover, the ensemble of atoms with dual I-AFC where the two frequency combs have similar mean frequency are capable of storing the polarization states of light efficiently. These two features together result in a quantum memory for vector-vortex beams. We discuss the factors which might affect the quality of the quantum memory and show that I-AFC based quantum memory for vector-vortex beams is robust. Further, we show that the Cs and Rb atoms serve as good candidates for storing vector-vortex states of light.

\begin{acknowledgments}
Chanchal acknowledges the Council of Scientific and Industrial Research (CSIR), Government of India, for financial support through a research fellowship (Award No. 09/947(0106)/2019-EMR-I). S.~K.~G.~ acknowledges  the financial support from Inter-disciplinary Cyber Physical Systems (ICPS) program of the Department of Science and Technology, India, (Grant No.:DST/ICPS/QuST/Theme-1/2019/12). We thank Louise Budzynski for useful discussions in the early stages of this project.
\end{acknowledgments}

\appendix

\section{Propagation of LG modes}\label{lg}
\subsection{In free space}
LG modes are the eigensolutions of the paraxial wave equation given by \cite{allen99,andrews2012}

\begin{align}
\qty[\laplacian_\perp+2\mi k\qty(\pdv{z}+\dfrac{1}{c}\pdv{t})]{\mathcal{E}}(\vb{r_{\perp}},z,t)=0,\label{Eq:paraxial}
\end{align}
where $\nabla^2_{\perp} \equiv \partial^2_x + \partial_y^2$.  Eq.~\eqref{Eq:paraxial} can be written in integral form by taking Fourier transforms in transverse position $(\vb{r_{\perp}}\to\vb{q})$, and the solution for the Fourier transformed electric field $\tilde{\mathcal{E}}(\vb{q},z,t)$ reads 
\begin{align}
\tilde{\mathcal{E}}(\vb{q},z,t)=\exp\qty({\dfrac{q^2}{2\mi k}z})\tilde{\mathcal{E}}\qty(\vb{q},0,t-\dfrac{z}{c}).
\end{align}
Now, taking inverse Fourier transform ($\vb{q} \to \vb{r}_{\perp}$), we get
\begin{align}
{\mathcal{E}}(\vb{r}_{\perp},z,t)=\mathcal{\mathcal{F}}^{-1}\qty[\exp\qty({\dfrac{q^2}{2\mi k}z})]*\mathcal{\mathcal{F}}^{-1}\qty[\tilde{\mathcal{E}}\qty(\vb{q},0,t-\dfrac{z}{c})],
\end{align}
where $*$ represents the convolution operation. Using the definition for convolution of two functions
\begin{align}
f(x)*g(x)=\int_{-\infty}^{\infty}f(x')g(x-x')~dx',
\end{align}
we can write the formal expression for ${\mathcal{E}}(\vb{r}_{\perp},z,t)$ as
\begin{align}
\mathcal{E}(\vb{r}_{\perp},z,t)=\int M(\vb{r_{\perp}}-\vb{r'_{\perp}},z)\mathcal{E}(\vb{r'_{\perp}},0,t-\dfrac{z}{c})d^2\vb{r'_{\perp}},\label{alg}
\end{align}
where
\begin{align}
M(\vb{r_{\perp}},z)=\mathcal{F}^{-1}\qty[\exp\qty({\dfrac{q^2}{2\mi k}z})]=\dfrac{1}{2\pi}\int e^{\mi \vb{q}.\vb{r_{\perp}}}e^{-\mi q^2z/{2k}}d^2\vb{q}.
\end{align}
Eq.~\eqref{alg} represents the electric field evolution in vacuum which does not affect the transverse profile of the field.

\subsection{Propagation of LG modes inside a medium}
The generalized paraxial wave equation inside a medium reads \cite{lukinlectures}
\begin{align}
\qty[\laplacian_\perp+2\mi k\qty(\pdv{z}+\dfrac{1}{c}\pdv{t})]{\mathcal{E}}(\vb{r},t)&=-\dfrac{k^2}{\epsilon_0}{\mathcal{P}}(\vb{r},t),\label{pe}
\end{align}
where $\mathcal{P}$ is the induced atomic polarization. 
Fourier transform of Eq.~(\ref{pe}) in the transverse plane and the time coordinate results in 


\begin{align}
\pdv{\tilde{{\mathcal{E}}}(\vb{q},z,\omega)}{z}=\dfrac{1}{2ik}\qty[(q^2+\dfrac{2wk}{c})\tilde{{\mathcal{E}}}(\vb{q},z,\omega)-\dfrac{w_Lk}{c \epsilon_0}\tilde{{\mathcal{P}}}(\vb{q},z,\omega)].\label{pw}
\end{align}
The atomic polarization amplitude ${\mathcal{P}}$ in terms of atomic coherence $\rho_{nm}$ between the atomic transition $\ket{n}\leftrightarrow \ket{m}$ can be written as~\cite{iafc19}
\begin{align}
{\mathcal{P}}(\vb{r}_{\perp},z,t)=2\mathcal{N}(\vb{r}_{\perp})\sum_{n,m}{d}_{nm}\rho_{nm}(\vb{r}_{\perp},z,t),\label{Pa}
\end{align}
where {$\mathcal{N}(\vb{r}_{\perp})$} is the atomic distribution function in the transverse plane. The same equation can be written upon taking the Fourier transform in the transverse plane and time, which reads
\begin{align}
\tilde{{\mathcal{P}}}(\vb{q},z,\omega)
=&2\tilde{\mathcal{N}}(\vb{q})*\sum_{n,m}{d}_{nm}\tilde{\rho}_{nm}(\vb{q},z,\omega).\label{Eq:P}
\end{align}

The dynamical equation for the atomic coherence corresponding to the Hamiltonian in Eq.~(\ref{hmlg}) can be written as
\begin{align}
  \begin{split}
\pdv{\rho_{nm}(\vb{r}_{\perp},z,t)}{t}+\Big(i&\Delta_{nm}+\dfrac{\gamma}{2}\Big)\rho_{nm}(\vb{r}_{\perp},z,t) \\
=&i{\dfrac{{d}_{nm}{{\mathcal{E}}(\vb{r}_{\perp},z,t)}}{2\hbar}}\rho_{mm}.\label{dy}
\end{split}
\end{align}


We can solve for $\rho_{nm}(\vb{q},z,w)$ by taking the Fourier transform  of Eq.~\eqref{dy} w.r.t  $t$ and $\vb{r}_{\perp}$. The expression for the Fourier transform of $\rho_{nm}(\vb{q},z,w)$ reads
\begin{align}
\tilde{\rho}_{nm}(\vb{q},z,w)&=\dfrac{i{d}_{nm}{\tilde{{\mathcal{E}}}(\vb{q},z,\omega)}\rho_{mm}}{2\hbar\qty[i(\Delta_{nm}+w)+\dfrac{\gamma}{2}]}.\label{ro}
\end{align}

Substituting Eq.~(\ref{ro}) in Eq.~\eqref{Eq:P} yields
\begin{align}
\tilde{{\mathcal{P}}}(\vb{q},z,\omega)=2\tilde{\mathcal{N}}(\vb{q})*\sum_{n,m}\dfrac{\mi {d}_{nm}^2{\tilde{{\mathcal{E}}}(\vb{q},z,\omega)}{\rho}_{mm}}{2\hbar\qty[\mi(\Delta_{nm}+w)+\dfrac{\gamma}{2}]}.\label{b}
\end{align}

Substituting above in Eq.~(\ref{pw}) gives

\begin{align}
\pdv{\tilde{\mathcal{E}}}{z}=&\qty(\dfrac{q^2}{2\mi k}-\dfrac{\mi\omega}{c})\tilde{\mathcal{E}}(\vb{q},z,\omega)-\tilde{\mathcal{N}}(\vb{q})*\mathcal{D}^{'}(\omega)\tilde{\mathcal{E}}(\vb{q},z,\omega),\label{Eq:ETilde}
\end{align}
where
\begin{align}
g'_{mm}=\dfrac{\omega_L\rho_{mm}}{2 c\hbar\epsilon_0},~
\mathcal{D}^{'}(\omega)=\sum_{nm}{\dfrac{g'_{mm}}{\qty[\mi(\Delta_{nm}+\omega)+\dfrac{\gamma}{2}]}d_{nm}^{2}}.
\end{align}
Solving Eq.~\eqref{Eq:ETilde} will yield the solution for the propagation of electric field through a medium. However, in general solving this equation is difficult.

A simple scenario is the homogeneous medium, where $\mathcal{N}$ is a constant. In this case, the equation can be simplified and the solution reads
\begin{align}
\tilde{\mathcal{E}}(\vb{q},z,\omega)=\exp\qty[\qty(\dfrac{q^2}{2\mi k}-\dfrac{\mi \omega}{c}-\mathcal{D}(\omega))z]\tilde{\mathcal{E}}(\vb{q},0,\omega).
\end{align}
where $\mathcal{D}(\omega)=N \mathcal{D}'(\omega)$ follows from Eq.~(\ref{pr})
Again, taking inverse Fourier transform ($w\to t$)
\begin{align}
\begin{aligned}
\tilde{\mathcal{E}}(\vb{q},z,t)=&\exp\qty(\dfrac{q^2z}{2ik})\mathcal{\mathcal{\mathcal{F}}}^{-1}\qty[\exp\qty(-\mathcal{D}(\omega)z)]\\
&*\mathcal{\mathcal{\mathcal{F}}}^{-1}\qty[\exp\qty(-\dfrac{\mi \omega z}{c})\tilde{\mathcal{E}}(\vb{q},0,\omega)],\\
\intertext{and applying convolution gives}
\tilde{\mathcal{E}}(\vb{q},z,t)=&\exp\qty(\dfrac{q^2z}{2ik})\qty[N(z,t))*\tilde{\mathcal{E}}(\vb{q},0,t-\dfrac{z}{c})],\\
=&\exp\qty(\dfrac{q^2z}{2\mi k})\qty[\int N(z,t-\tau)\tilde{\mathcal{E}}(\vb{q},0,\tau-\dfrac{z}{c})d\tau],
\end{aligned}
\end{align}
where,
\begin{align}
N(t,z)=\mathcal{\mathcal{\mathcal{F}}}^{-1}\qty[\exp\qty(-\mathcal{D}(\omega)z)]=\dfrac{1}{2\pi}\int e^{\mi \omega t}e^{-\mathcal{D}(\omega)z}dw.
\end{align}
Now, taking inverse Fourier transform ($\vb{q} \to \vb{r}_{\perp}$) gives
\begin{equation}
\begin{split}
{\mathcal{E}}(\vb{r}_{\perp},z,t)=&\mathcal{F}^{-1}\qty[\exp\qty(\dfrac{q^2z}{2\mi k})]\\&*\mathcal{F}^{-1}\qty[\int N(t-\tau,z)\tilde{\mathcal{E}}(\vb{q},0,\tau-\dfrac{z}{c})d\tau],\\
=&M(\vb{r_{\perp}},z)*\qty[\int N(t-\tau,z){\mathcal{E}}(\vb{r_{\perp}},0,\tau-\dfrac{z}{c}))d\tau],\\
=&\int M(\vb{r_{\perp}}-\vb{r'_{\perp}},z)\\&\qty[\int N(t-\tau,z){\mathcal{E}}(\vb{r'_{\perp}},0,\tau-\dfrac{z}{c})d\tau] d^2\vb{r'_{\perp}}.
\end{split}
\end{equation}

%

\end{document}